\documentclass[aps,superscriptaddress,twocolumn,showpacs,preprintnumbers,amsmath,amssymb]{revtex4}
\usepackage[cp1251]{inputenc}
\usepackage{amssymb,amsmath}
\usepackage[mathscr]{eucal}
\usepackage{graphicx}
\usepackage{longtable}
\usepackage[dvips]{hyperref} 
\begin{document}
\title{Spin-Peierls transition in carbynoid conductors: infrared absorption study}

\author{D.Yearchuck}
\affiliation{Minsk State Higher College, Uborevich Str.77, Minsk, 220096, RB; dpy@tut.by}
\author{E.Yerchak}
\affiliation{Belarusian State University, Nezavisimosti Ave.4, Minsk, 220030, RB; Jarchak@gmail.com}

\date{\today}
             
\begin{abstract}
The results of IR-studies in quasi-1D carbynoid films produced by dehydrohalogenation of poly(vinilidene fluoride) are in good agreement with the assumption that carbynoid films studied are generalized spin -Peierls conductors, the metal to insulator transition in which can be described in the frame of t-J model. Residual atoms of fluorine, hydrogen and atoms of main technological impurity oxygen in the form of various complexes in interchain space are suggested to be spin - (or joint spin - and electrical) conductivity dopants.
Antiferroelectric spin wave resonance (AFESWR) being to be optical analogue of antiferromagnetic spin wave resonance has been identified for the first time. Electric spin-Peierls polaron lattice in C-C -bonds is proposed to be responsible for the observed AFESWR both in starting PWDF films and in carbynoid B-films (the samples with the least impurity content). Electric spin moment with pure imaginary value predicted by Dirac as early as 1928 was identified for the first time. Electric spin-Peierls polarons are proposed to be electric spin moment carriers. 
It has been established that topological solitons, earlier called spin-Peierls solitons (SPS), are simultaneously active, unlike to topological solitons with nonzero spin in \textit{trans}-polyacetylene, in both optical and magnetic resonance spectra.It is explained in suggestion that SPS possess by both electric and magnetic spin moments which can be considered as two components of complex electromagnetic spin vector as a single whole. SPS proposed to be consisting of two coupled domain walls in both magnetic and electric generalized spin density wave (GSDW), produced by electromagnetic spin-Peierls transition in its generalized form in $\pi$ - and $\sigma$ -subsystems of carbynoids. 
\end{abstract}

\pacs{78.30.-j, 76.30.-v, 76.50.+g, 78.67.-n}
\maketitle

\section{INTRODUCTION AND BACKGROUND}

The main feature of carbynoids which distinguishes advantageously carbynoids 
from all known organic conductors including the most studied organic 
conductor \textit{trans}-polyacetylene (\textit{t}-PA) as well as intensively studied at present 
carbon nanotubes and fullerenes is availability of two $\pi$-electronic 
subsystems. This feature along with the possibility to produce the materials 
with various kink size in their chain-kinked structure open a multiplicity 
of very interesting physical properties. So, ESR studies of carbynoid films 
produced by dehydrofluorination of poly(vinilidene fluoride) have shown that 
they can be described by quasi-1D spin-Peierls conductor model with spin 
-Peierls transition in generalized form \cite{Ertchak_Carbyne_and_Carbynoid_Structures}. It is understandable that $\pi$-electronic subsystems have to be in carbynoids rather strongly 
correlated. Consequently, spin-Peierls transition has to be accompanied by 
spin-charge separation effect in accordance with general guantum field 
analysis of 1D strongly correlated electronic systems, see for instance \cite{Mudry_Fradkin}. 
Note that spin-charge separation effect was established in 1D conductors 
for the first time in pioneering works of Su, Schrieffer, and Heeger (SSH) 
\cite{Su_Schrieffer_Heeger_1979, Su_Schrieffer_Heeger_1980}. Su, Schrieffer, Heeger identified by analysis of, mainly, ESR and IR 
absorption data on the base of theoretical model, which is well known at 
present as SSH-model, the existence in \textit{trans}-polyacetylene of topological 
solitons, which carry electronic spins without electric charge and \textit{vice versa}. 
Topological solitons in \textit{t}-PA as it is argued in \cite{Ertchak_Physica_Status_Solidi} produce a class in 
soliton family, SSH-solitons. Their properties in \textit{t}-PA and consequently spin 
-charge separation effect which accompanies in \textit{t}-PA the usual \textit{Peierls} transition 
are studied rather well. At the same time the effect of spin-charge 
separation did not studied (to our knowledge) in spin-Peierls conductors 
although there are to be known at present a number of spin-Peierls 
conductors. There is some progress in this direction the only in carbynoids. 
Really a mechanism of spin-charge separation in 1D strongly correlated 
electronic systems has to be topological in its nature and has to be 
realized by topological solitons' formation \cite{Mudry_Fradkin}. Topological solitons were identified in carbynoids by ESR-study ( C-M2 spectrum) \cite{Ertchak_Carbyne_and_Carbynoid_Structures,Ertchak_J_Physics_Condensed_Matter}. From mathematical point of view they 
can be referred to SSH class, but physically they represent a new kind of solitons. These quasiparticles, according to \cite{Ertchak_Carbyne_and_Carbynoid_Structures,Ertchak_J_Physics_Condensed_Matter} are domain walls in dimerised \textit{spin density 
distribution} in distinction from SSH-solitons in \textit{t}-PA, which are domain walls in 
dimerised \textit{bond patterns}. Topological solitons in carbynoids were called spin-Peierls 
solitons (SPS). The view on the nature of SPS will be developed in this paper.
 Especially interesting that preliminary results of IR 
absorption study, reported very briefly in \cite{Kudryavtsev_Yearchuck, Yearchuck_Strasbourg_2004, Yearchuck_Wiesbaden_2004}, have been shown, that 
spin-Peierls solitons simultaneously with ESR activity are optically active. 
It means that mechanism of spin-charge separation has the peculiarities in 
carbynoids. The aim of presented work is to understand this phenomenon and 
therefore to grasp in mechanism of spin-charge separation as well as in 
related phenomena accompanying spin-Peierls transition.

It should take in mind some background details, which can help to achieve 
the aim. They are as follows. The magnetic spin wave patterns, obeying to k$^{2}$ 
-dispersion law (like to that one observed in ferromagnetic thin films) and 
indicating on new mechanism of room temperature (RT) ferromagnetic ordering 
of carbon atoms in carbynoid chains, have been observed. Carbynoid films 
seem to be therefore the first substance among free organic conductors at 
all with RT-ferromagnetic ordering. Note that ratio of magnetic spin 
exchange constants J in carbynoids \cite{Ertchak_J_Physics_Condensed_Matter} and in cobalt films obtained by 
comparison the results presented in \cite{Ertchak_J_Physics_Condensed_Matter} and \cite{Tannenwald_Weber_1961} is about 0.63 and 0.76 for 
two magnetic spin wave resonance patterns registered in carbynoid samples in 
1.8 Y after their production. Here the values of lattice parameter $a$, equaled 
to 0.128 nm and 0.3537 nm, were used for carbynoids and cobalt 
correspondingly. It allows to evaluate the Curie temperature (suggested to 
be simple average of values corresponding to two ratios of exchange 
constants J) $\sim$  777$^{o}$C. It is interesting that stability of 
ferromagnetic ordering in carbynoids is much higher in comparison with the 
stability of all other magnetically ordered organic conductors. For instance 
in known organic antiferromagnetically ordered spin density wave (SDW) 
conductors (DMET, MDTTF, TMTSF-salts, etc) SDW-transition takes place at 
$\sim$ 20K or even lower \cite{Kikuchi,Nakamura,Jerome}. Note that ordering in carbynoids is 
formed in outer shell in distinction from that one produced by unpaired 
spins in inner atomic shells of transition metal or rare earth elements. 
Consequently the properties of ordered state in both cases should be quite 
different.

\begin{figure}[t]
\includegraphics[width=0.5\textwidth]{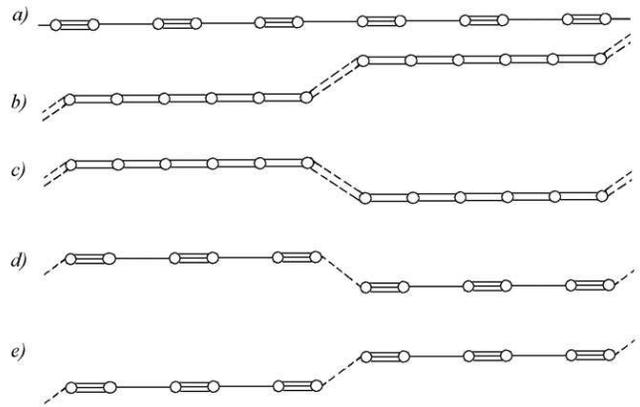}
\caption[Some Peierls transition schemes for carbyne chain]{\label{fig1} Some Peierls transition schemes for carbyne chain: a) Peierls transition accompanying by bond length dimerization; b) generalized Peierls transition accompanying by kink bond angle alternation (\textit{trans}-configuration); c) generalized Peierls transition accompanying by kink bond angle alternation (\textit{cis}-configuration); d) mixed state in result of simultaneous Peierls transition accompanying by bond length dimerization and generalized Peierls transition accompanying by kink bond angle alternation (\textit{cis}-configuration); e) mixed state in result of simultaneous Peierls transition accompanying by bond length dimerization and generalized Peierls transition accompanying by kink bond angle alternation (\textit{trans}-configuration).}
\end{figure}

Second detail is concerned the generalization of both Peierls and 
spin-Peierls transitions. It should be noted that representation of pure 
carbynes as Peierls systems, was established in fact by Heimann \textit{et al} in \cite{Heimann1983,Heimann1984}, where it was proposed that the starting linear atomic chain with 
equidistant interatomic distance becomes kinked (in fact angle and/or bond 
dimerized) with C$_{n}$-linear$ n$-atomic fragments (kinks), $n$ equal to 6 - 12, being 
to be dimerazation units, Fig.\ref{fig1}. 

This idea was developed on a language of Peierls transition in generalized 
form in \cite{Ertchak_Carbyne_and_Carbynoid_Structures}. The kink size, interkink angles and, consequently, resulting 
electronic and geometrical chain structure are determined by the value of 
electron-electron interatomic interaction along chain, which in its turn is 
dependent, to definite extent, on production conditions. This circumstance 
means, that physical properties of carbynes and carbynoids can be varied in 
very wide range. It seems to be substantial for their practical use. 
It is 
interesting to try to predict the possible carbon structures, for which 
Peierls transition should also be realized in its generalized form. Quasi 
-1D-carbon clusters, `` claimed'' to be dimerization units, have to be 
evidently stable. The linear carbon clusters are the most stable up to 
C$_{12}$ \cite{Weltner1989} in a full accordance with carbynes' structure \cite{Weltner1989}. From 
C$_{13}$ to C$_{25}$ the monocyclic ring clusters are preferable in the 
absence of the terminal groups \cite{Weltner1989}, although the monocyclic ring clusters 
can exist according to \cite{Helden1993} from $n = $7 to $n = $40. Note, however, that in the stability interval of
monocyclic rings the linear isomers with 8 to 28 atoms 
were also synthesized by the addition of nonreactive terminal groups \cite{Lagow1995}. 
Starting from $n = $21 the polycycle structures are possible \cite{Feyereisen1992} and starting 
from $n = $32 the fullerene structures were found to be the most stable \cite{Feyereisen1992,Jones1997}. 
Consequently all listed above stable clusters can be ``claimed'' in the capacity of dimerization units of corresponding 1D-carbon materials. For instance, if 
quasi-1D carbon chain is formed from the carbon units with number of atoms $n $equaled 32 or more pro unit the preferable geometrical 
structure for the separate unit should be fullerene structure. Indeed in 
accordance with experimental data in \cite{Zhu_Cox_Fisher_1995} the second known quasi-1D system, 
where the generalized Peierls transition was experimentally observed, is the 
quasi-1D-phase of KC$_{60}$ in which the C$_{60}$ ions form chains with 
alternating longer and shorter center to center separation. Center to center 
separation between KC$_{60}$-dimers according to \cite{Zhu_Cox_Fisher_1995} is 9.34 A\r{ }. 

\textit{Therefore it reasonably to suggest that the dimerisation units with more 
than 12 atoms by generalized Peierls transition in carbon quasi-1D chains 
will possess by nonlinear geometric structure. }

The paper is organized as follows. In Sec.II the experimental technique and 
some details of ESR- and IR-measurements are described. In Sec.III the 
experimental data of IR studies in carbynoids are presented. The data of IR 
studies are compared with ESR-data. It is shown that IR- and ESR-data 
correlate well and can be explained in the frames of the same model. In 
Sec.IV the summary and conclusions are represented.

\section{Experimental Technique }

IR absorption studies have been fulfilled on unoriented and uniaxially 
oriented carbynoid film samples prepared by chemical dehydrohalogenation of 
poly(vinylidene fluoride) (PVDF). Preparation details were described in \cite{Ertchak_J_Physics_Condensed_Matter}. 
Due to strong absorption in the most interesting spectral range IR 
-measurements have been done repeatedly on the same samples but grinded and 
compressed into pellet with KBr. IR measurements have also been done for 
comparison on starting unoriented and uniaxially oriented PVDF-films as 
well as on grinded PVDF-films and compressed into pellet with KBr. Uniaxial 
orientation of PVDF-films was provided by the procedure which was identical to the 
procedure used by the production of oriented carbynoid films.
 Among the IR studied sets of samples there were 
two sets (designated A and B) which have been earlier studied by electron spin resonance 
(ESR) \cite{Ertchak_J_Physics_Condensed_Matter}. We report now infrared (IR)-study results 
in comparison with ESR results that is those ones obtained the only on these 
two sets of samples. Carbynoid samples contained a rather high concentration of 
residual fluorine and technological oxygen atoms. The samples of A set 
(hereinafter A-samples, they were designated as the second series samples in 
 \cite{Ertchak_J_Physics_Condensed_Matter}) were with F/C ratio equal to 3/7, their oxygen contamination O/C was 1 to 5. A-samples were thermally treated at 120\r{ }C for 2 hours. The contamination of fluorine and oxygen atoms in the samples of B series 
(hereinafter B-samples, they were designated as third series samples in 
 \cite{Ertchak_J_Physics_Condensed_Matter}) was intermediate between 3/10 and 3/7 for the F/C ratio and between 1 
to 10 and 1 to 5 for O/C ratio (however O and F content 
was not determined exactly). Like to the classification proposed for 
doped $t $-PA \cite{Heeger_1988}, the samples studied can be attributed both to doped carbynes 
and to carbynoids that is to materials including a wide range of carbyne-like structures. 

\section{RESULTS AND DISCUSSION}

\subsection{IR spectra in carbynoid samples}

\subsubsection{Film samples}

\begin{figure}[t]
\includegraphics[width=0.5\textwidth]{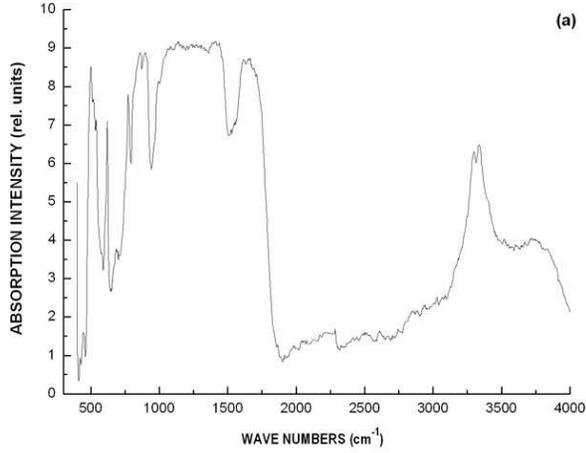}
\includegraphics[width=0.5\textwidth]{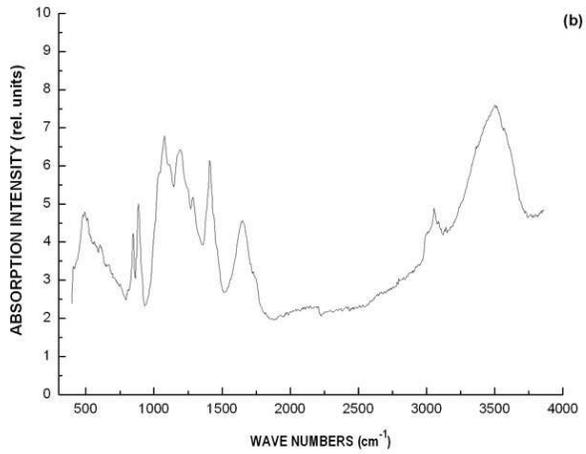}
\caption[Spectral distribution of IR absorption intensity in carbynoid B-film sample]{\label{fig2} Spectral distribution of IR absorption intensity in carbynoid B-film sample (a) top picture: uniaxially oriented carbynoid film, (b) bottom picture: B-film was grinded and compressed into pellet with KBr.}
\end{figure}

\begin{figure}
\includegraphics[width=0.5\textwidth]{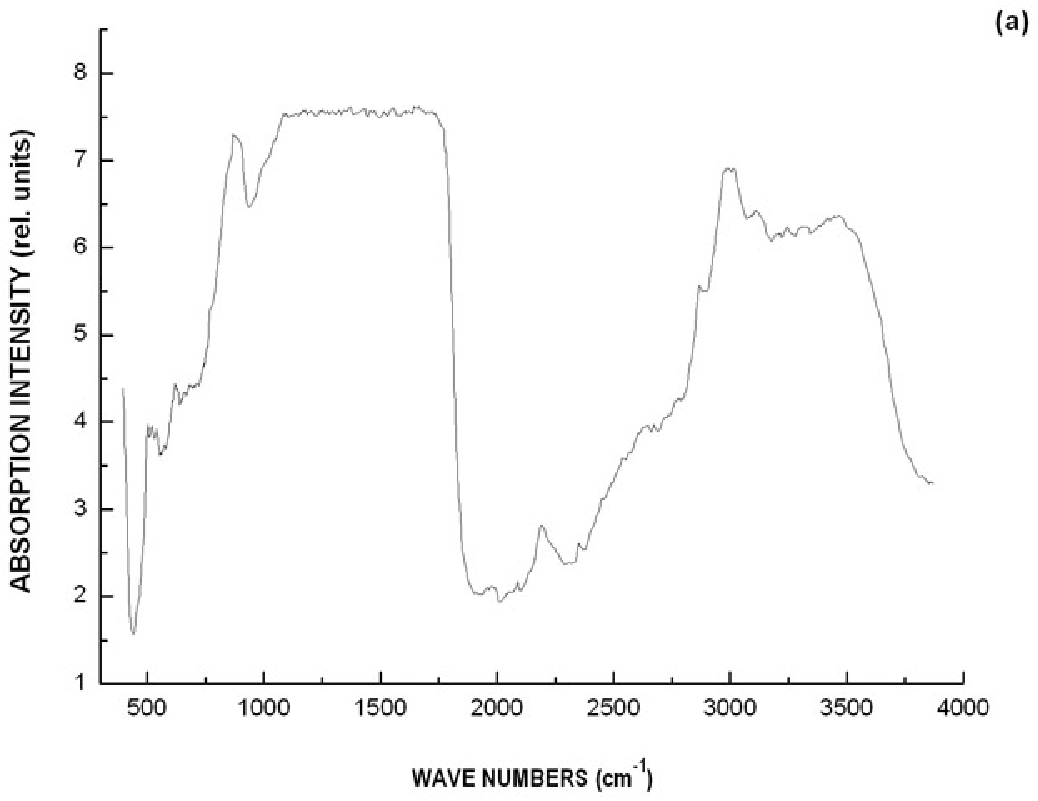}
\includegraphics[width=0.5\textwidth]{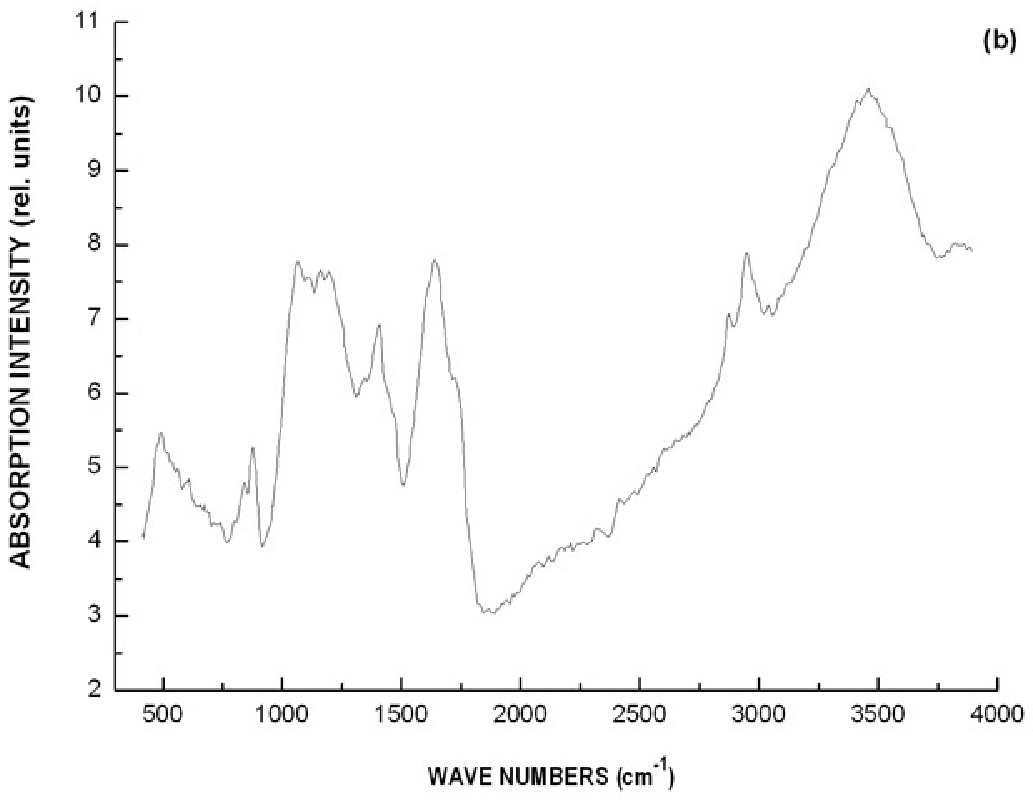}
\caption[Spectral distribution of IR-absorption intensity in carbynoid A- film sample] {\label{fig3} Spectral distribution of IR-absorption intensity in carbynoid A-film sample (a) top picture: uniaxially oriented carbynoid film, (b) bottom picture: A-film was grinded and compressed into pellet with KBr.}
\end{figure}

\begin{figure}
\includegraphics[width=0.51\textwidth]{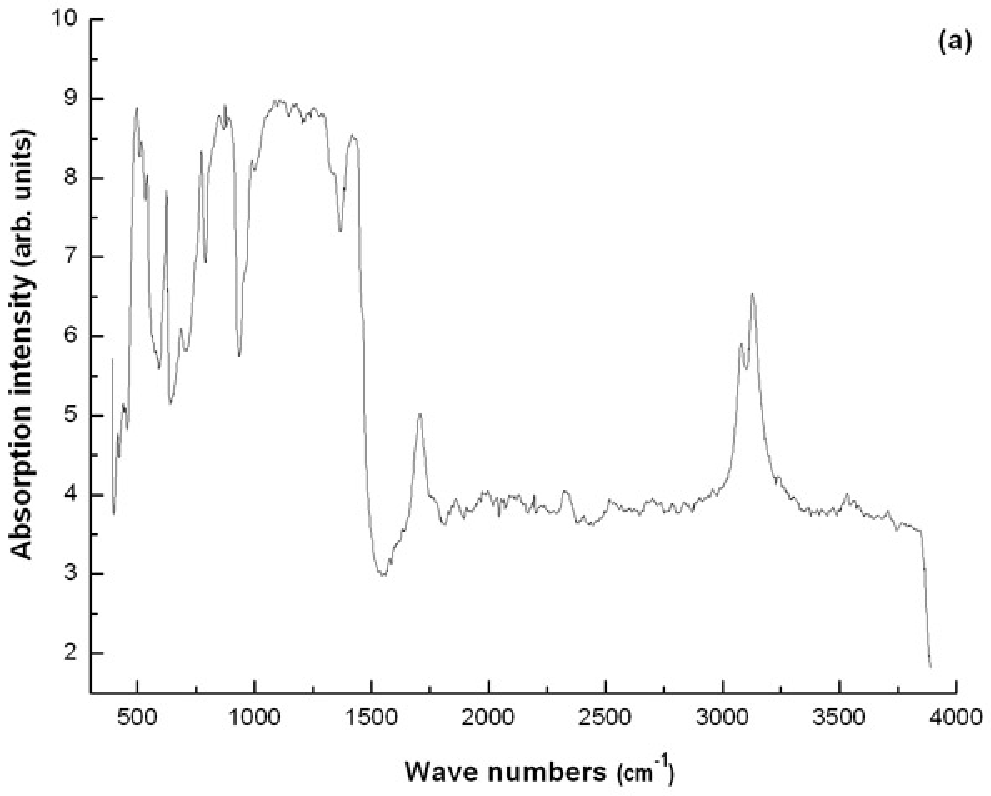}
\includegraphics[width=0.51\textwidth]{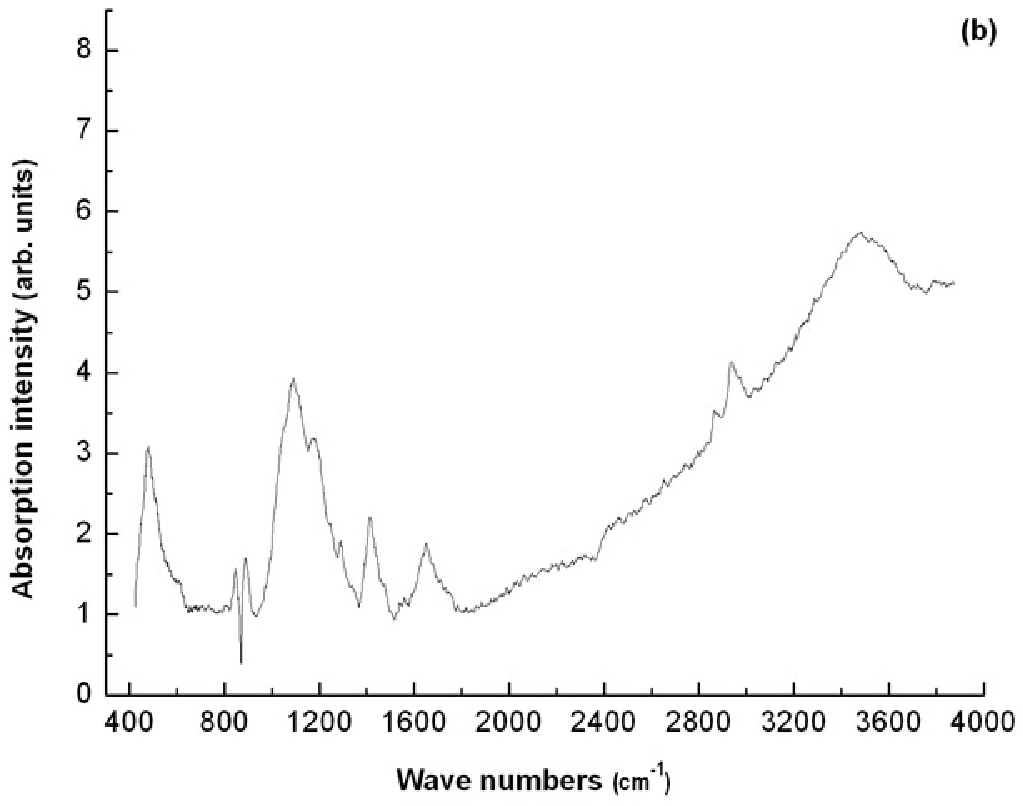}
\caption[General view of spectral distribution of IR absorption intensity in PVDF]{\label{fig4} General view of spectral distribution of IR absorption intensity in PVDF: (a) top picture: uniaxially oriented PVDF-film, (b) bottom picture: PVDF-film was grinded and compressed into pellet with KBr.}
\end{figure}

Difference between the spectra which belong to the samples of the same sets 
was negligible. Therefore data the only on two (A and B) unoriented, two (A 
and B) oriented and two (A and B) grinded samples are presented. General 
view of IR spectra, registered on oriented carbynoid A- and B-films and on 
the same samples, but being to be grinded and compressed into pellet with KBr after measurements in film form, is presented in Figures \ref{fig2}, \ref{fig3} correspondingly. General view of IR spectra, registered subsequently on 
oriented PVDF-film and then in the same sample, being to be grinded and 
compressed into pellet with KBr, is presented in Figure \ref{fig4}. Detailed data, concerning the positions of observed absorption bands and their relative amplitudes are summarized in Tables \ref{table1}, \ref{table2}. Main difference of IR spectra of carbynoid films from IR spectra of starting PVDF-films is, as it is seen from Figures,the appearance of IR-activity in the range near 1600 - 1800 cm$^{-1}$ and the emergence at higher frequencies of broad asymmetric line (BAL) extending in rather wide range for both oriented and unoriented samples (for instance, in the range of 2000 to 3760 cm$^{-1}$ in unoriented B-sample). Maximum positions of BAL are equal to $\sim $ 3467 and $\sim $ 3425 cm$^{-1}$ for A- and B-samples correspondingly. Qualitatively the same picture was observed in a number of carbynoids produced by other methods, see for review, e.g. \cite{Evsyukov1999}. 
Especially interesting is that the asymmetry character of BAL, that is, the 
ratio of effective values of extension of left and right parts of intensity 
distribution curve relatively the maximal value, is being to be opposite in 
comparison with that one registered for analogous broad line, which was 
observed and well studied in \textit{trans}-polyacetylene \cite{Blanchet1983}. Effective values of BAL width and relative amplitudes are $\Delta \nu ^{A}_{BAL} \approx 840 cm^{-1}$, $a_{BAL} = 0.60$ and $\Delta \nu ^{{\rm B}}_{BAL}\approx 820 cm^{-1}$, $a_{BAL} = 0.32$ in A and B-films correspondingly. Therefore \textit{the width of BAL and its relative amplitude are increasing in the sequence of B- to A-films. Given increase seems to be correlating with the total impurity increase.}

\textit{There is also essential increase, equal to 1.8, of relative amplitude of the line with position near 2200 cm}$^{-1}$\textit{ in the sequence of B- to A-films.} It is interesting that this ratio is almost coinciding with 
corresponding ratio of relative amplitudes for BAL, which is equal $\sim 
$1.9. The line near 2200 cm$^{-1}$ is characteristic line for polyyne type 
of carbyne and carbynoid electronic structure \cite{Evsyukov1999}. Note that \textit{there is some red shift from 2210 cm}$^{-1}$\textit{ in B-film to 2195 cm}$^{-1}$\textit{ in A-film for the position} of 
the band indicated. Note also that the band near 2200 cm$^{-1}$ is 
relatively week pronounced in the samples obtained by dehydrohalogenation of 
PVDF in comparison with those ones obtained for instance by 
dehydrohalogenation of poly(vinilidenchlorides) in the agreement with 
previous observations described in \cite{Sladkov1989}. 
\begin{figure}
\includegraphics[width=0.52\textwidth]{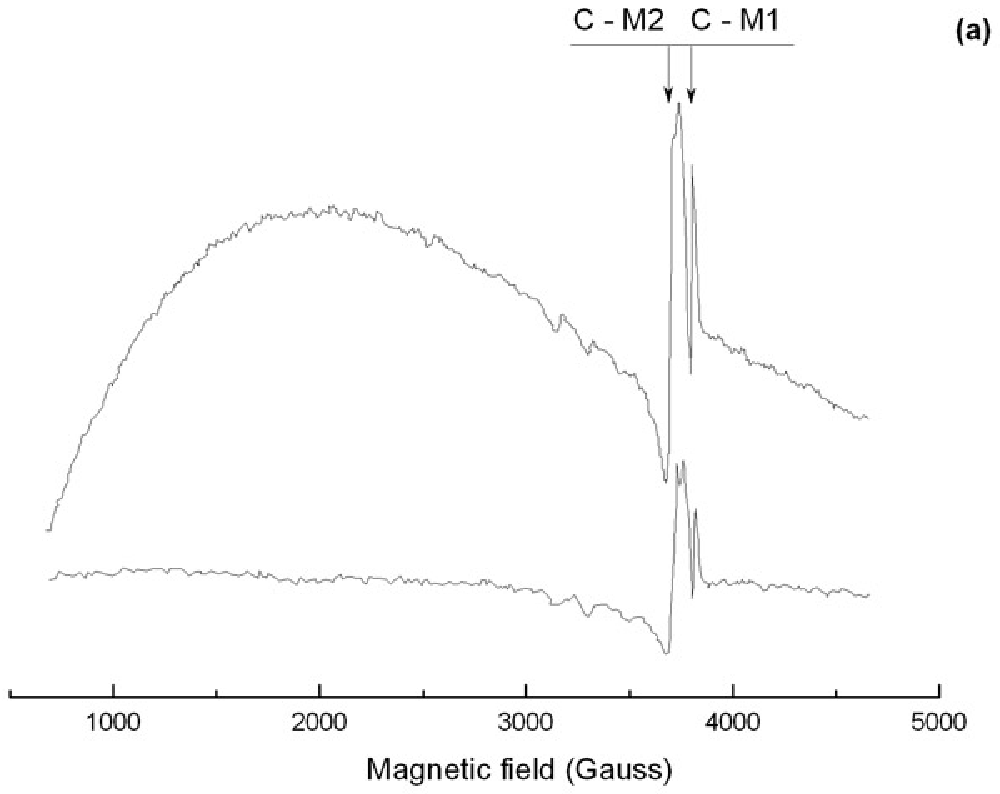}
\includegraphics[width=0.485\textwidth]{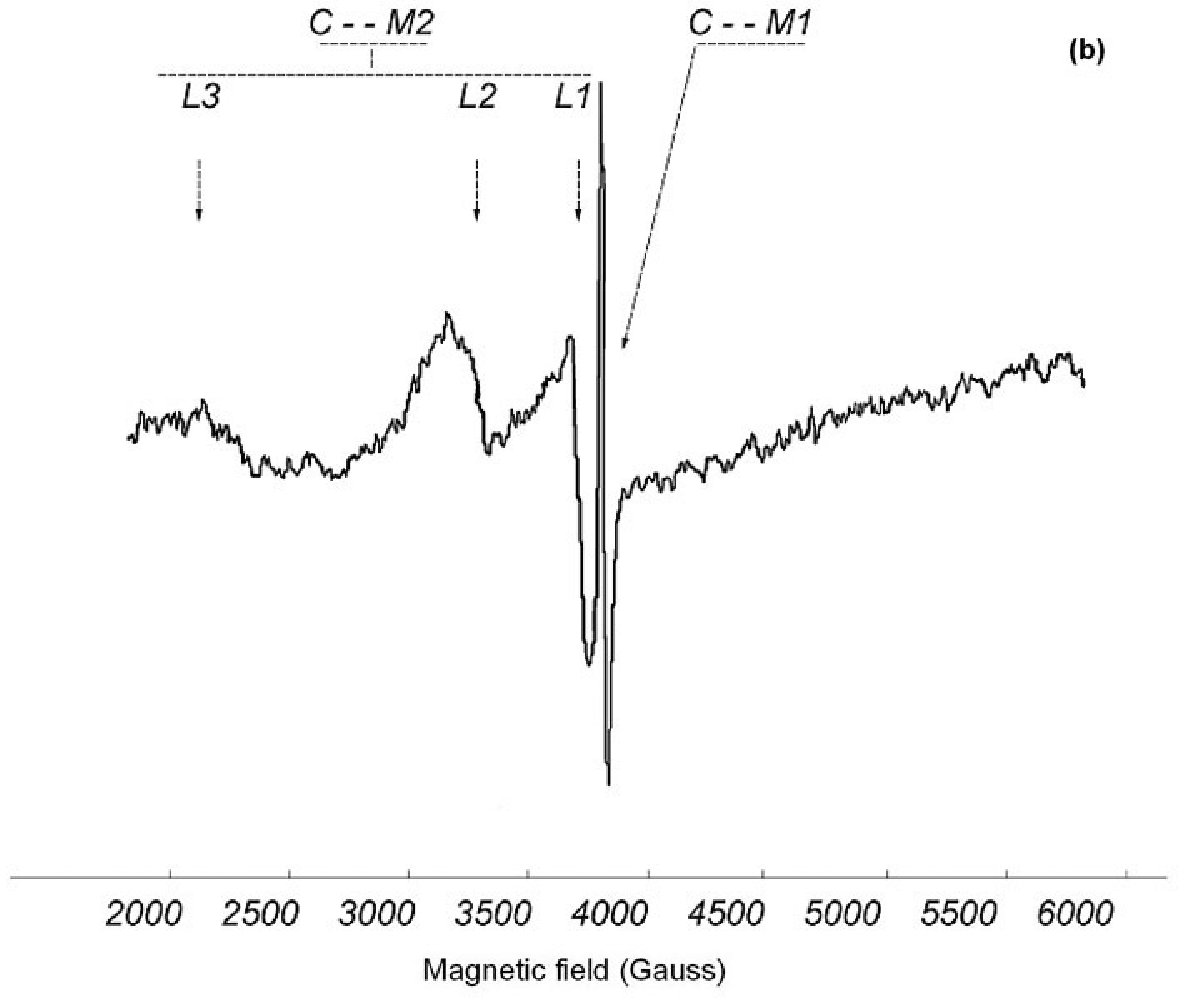}
\caption{\label{fig5} (a)Top picture: ESR-absorption immediately after sample 
preparation (top spectrum) and in 12 days (bottom spectrum) in an oriented 
carbynoid B-film sample. The intensity scale is identical for 
top and bottom spectra. (b) Bottom picture: ESR-absorption in an oriented 
carbynoid A-film sample. The spectra are reproduced from \cite{Ertchak_Carbyne_and_Carbynoid_Structures} 
and \cite{Ertchak_J_Physics_Condensed_Matter}}
\end{figure}

\begin{figure}
\includegraphics[width=0.5\textwidth]{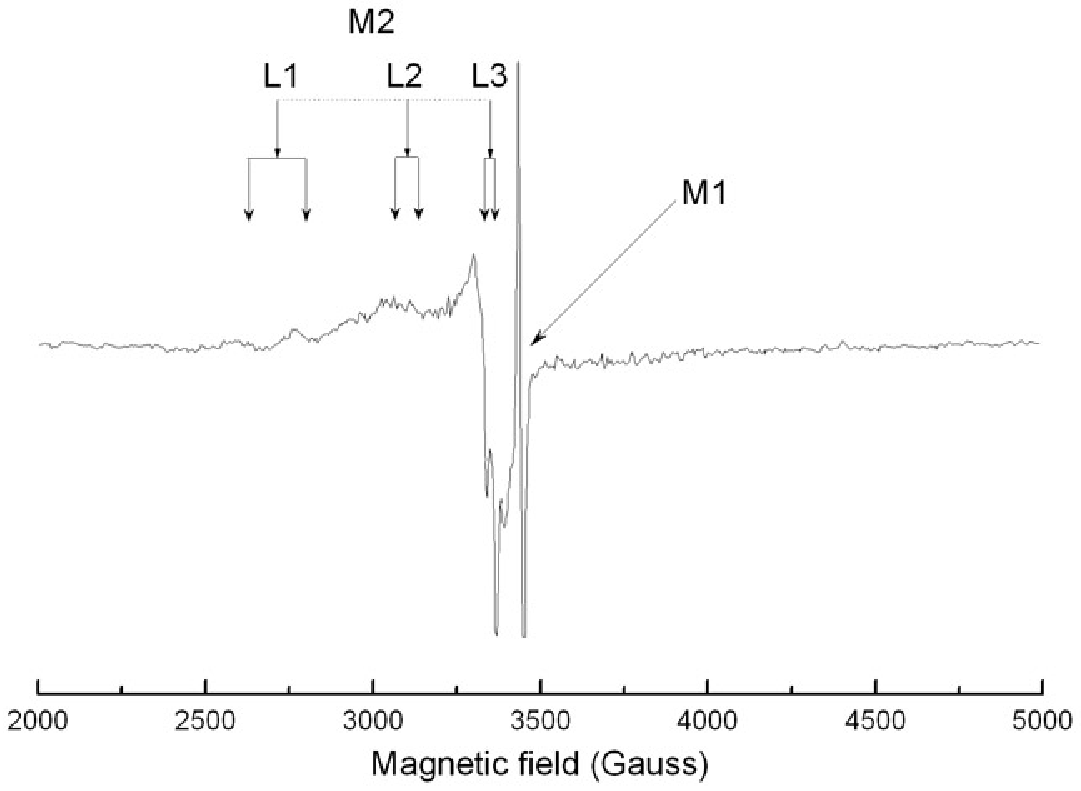}
\caption[ESR-spectrum in A-film sample, measured in 1.8 
year after sample preparation.]{\label{fig6} ESR-spectrum in A-film sample, measured in 1.8 
year after sample preparation. The sample plane is perpendicular to magnetic 
field vector of microwave field, the static magnetic field is parallel to 
the chain orientation axis, resonance frequency $\nu $ = 9644.70 MHz. The spectrum is reproduced 
from \cite{Ertchak_J_Physics_Condensed_Matter}.}
\end{figure}
The structure of IR-absorption bands in the range 400 - 1000 cm$^{-1}$ is 
almost coinciding in both oriented and unoriented B-film samples, Table 
\ref{table1}. Especially interesting that absorption structure is very similar to that 
one observed in starting PVDF film, compare Fig.\ref{fig2}, Fig.\ref{fig4}. The absorption in 
the range 400 - 590 cm$^{-1}$ represents itself the single transition band with 
fine structure. Absorption counter has a resemblance to absorption 
derivative shape or dispersion shape, although it is asymmetric to some 
extent. It is especially clear seen in Fig.\ref{fig4} for PVDF sample. If it so, then one should consider dip positions in the range 400 - 470 cm$^{-1}$ to be 
signal peaks. In this suggestion the observed fine structured band becomes clear interpretation. It is typical vibration-rotation band corresponding to vibration transition at 475, 475.5, 477 cm$^{-1}$ in PVDF-, B-unoriented, B-oriented films correspondingly. Really, peaks at 495, 517, 538 cm$^{-1}$, shoulder at 557 cm$^{-1}$, weak peak at 577 cm$^{-1}$ in B-unoriented film, peaks at 497, 517, 538 cm$^{-1}$, shoulder at 555 cm$^{-1}$, weakly pronounced peak in the range 575 - 580 cm$^{-1}$ in B-oriented film and peaks at 495, 514, 537 cm$^{-1}$, shoulder in the range 555 - 560 cm$^{-1}$, weakly pronounced peak at 575 cm$^{-1}$ in 
PVDF-film are almost equidistant lines with subsequent addition of $\sim $20 
cm$^{-1}$ to the line center at 475 - 477 cm$^{-1}$, that is all these 5 
lines represents well known R-branch. To P-branch one should refer the lines 
with positions at 456, 423 cm$^{-1}$ in B-unoriented film, 457, 424 
cm$^{-1}$ in B-oriented film and the lines at 455, 423, 398 cm$^{-1}$ in 
PVDF-film. The rotation transition at the center of band, that is at 475 --
477 cm$^{-1}$, is absent (as usually). Note that the observation of P-branch at 
dip positions in its turn testifies the correctness of the conclusion that \textit{signal from the absorption centers which are responsible for the fine-structured band is registered in dispersion mode}. The nature of this band will be discussed in Sec.G. It is essentially that the structure of this band is almost vanishing in A-films, the only very weak peaks at 500, 520, 540 cm$^{-1}$ are visible on the noise level. Amplitude decrease 
relatively to off-scale value is $\sim $2.3. The width of this band is also 
decreasing from $\sim $100 cm$^{-1}$ in PVDF and B-carbynoids to $\sim $60 
cm$^{-1}$ in A-carbynoids. Taking into account slightly different off-scale 
values (scaling factor is 1.3) and the shape of registered signal we obtain then for intensity decrease the value of 8.3. It is also substantial that shape and intensity of the band near 3000 cm$^{-1}$ with peaks at 3038, 2998 cm$^{-1}$ and dip at 3013 cm$^{-1}$, 
registered in B-sample, are also changed drastically in A-film, compare Fig.\ref{fig2} 
and Fig.\ref{fig3}. Evaluations, which were carried out analogously to those ones for fine-structured band, give the factor $\sim $8.6 for intensity 
decrease of the band near 3000 cm$^{-1 }$ in A-sample. The band near 3000 
cm$^{-1}$ was also observed in starting PVDF-film with the same dip position 
at 3013 cm$^{-1}$ like to that one in B-sample. At the same time the band 
has two peaks at 2992, 3033 in oriented PVDF-sample and two peaks at 2990, 
3034 cm$^{-1}$, in PVDF-unoriented sample, that is, they are at different positions, 
shifted to low frequencies, in comparison with those ones in carbynoid 
B-sample. It seems to be natural to attribute the band near 3000 cm$^{-1}$, 
registered both in B-sample and in PVDF to C-H vibration mode. However, 
observed shift of band peaks seems to be indicating on different surrounding 
of vibration centers. The result agrees well with the suggestion that C-H bonds in 
carbynoid samples belong to some interchain complexes mainly, but 
in PVDF-sample the band near 3000 cm$^{-1}$ is vibration mode of lateral valence C-H bonds in chain structure. The shift to low frequencies in PVDF is explained in a natural 
way, being to be well known shift, taking place by polymerization process 
including the same C-H units, see for instance \cite{Gribov1977}. [It can be shown that 
the same conclusion follows from shape analysis of the band near 3000 cm$^{-1}$ 
both in PVDF and in carbynoid B-sample]. In favour of this suggestion can 
also indicate the presence of many shoulders near 3000 cm$^{-1}$, that is, 
shoulders at 3190, 3140, 3108, 3090, 2985, 2950, 2850 cm$^{-1}$ which are 
detected in oriented B-sample, but they are absent in both oriented and 
unoriented PVDF films. Consequently 
the above listed shoulders cannot belong to C-H lateral bonds of polymer carbynoid
chain. Especially interesting that shoulders near 3108 and 
at 2950 cm$^{-1}$ are also presenting in unoriented B-sample, but they are 
less pronounced and the first shoulder has slightly different frequency, 3100 
cm$^{-1}$. At the same time the shoulders at 3190, 3140, 3090, 2985, 2850 cm$^{-1}$ as well as the shoulder at 3538 cm$^{-1}$ are also absent in unoriented B-film. The situation is very similar to averaging-out of spectra for various configurations of nonisotropic point centers in 
any single crystal by its grinding. The role of such point centers can play the 
simple molecules which have finite discrete set of possible orientation of C-H bonds in 
interchain space of uniaxially ordered film with slightly different resonance frequencies. Therefore it seems to be reasonable to ascribe the observed shoulders, observed in oriented carbynoid B-sample, along with main two peaks near 3000 cm$^{-1}$ to C-H bonds in simple 
molecules or molecular complexes which include C-H bond or several C-H bonds and which are localized in interchain space in some set of energetically slightly different configurations. 
Note, that the decrease of C-H lateral bonds corresponds by dehydrohalogenation process to 
equal decrease of C-F lateral bonds \cite{Evsyukov1999}. Then we obtain in fact the proof 
that \textit{the number of hydrogen and fluorine atoms, which are considered to be responsible in their lateral to carbon backbone valence C-H and C-F configurations is much less (if they are existing at all) in carbynoid B-sample in comparison with those ones in PVDF-sample.} It is interesting that in the films of A-set even the number of IR-active complexes with C-H vibration mode is also substantially less (by foregoing factor $\sim $8.6) in comparison with those ones in B-set.
Thus \textit{hydrogen, fluorine (and evidently oxygen) atoms form more 
effectively the interchain molecules and/or complexes in both B- and A-samples. There is clear tendence to decrease of concentration of optically active complexes (at 
least of those ones including C-H-bonds) in B- to A-sample sequence.} \begin{widetext}
\setlongtables 
\begin{longtable}
{||p{112pt}|p{45pt}||p{112pt}|p{45pt}||p{112pt}|p{45pt}||}
\caption{\label{table1}Spectral positions and amplitudes of the lines registered in carbynoid film samples. 
Here wpp means ``weakly pronounced peak'', designations (p) $\leftrightarrow$ (d) and (p) $\leftrightarrow $ (p) are differences between amplitudes of  
concrete peak and neighbour dip or peak correspondingly. Amplitude was determined relatively to off-scale value between 1750--1600 cm$^{-1}$.}\\ \hline 
\multicolumn{2}{||c||}{\textit{Unoriented B-sample}}&\multicolumn{2}{c||}{\textit{Oriented B-sample}}&\multicolumn{2}{c||}{\textit{Oriented A-sample}}  \\ \hline
Frequency, cm$^{-1}$& Amplitude& Frequency, cm$^{-1}$& Amplitude& Frequency, cm$^{-1}$& Amplitude \\ \hline
$\sim 3425,$ BAL max.&0.32&$\sim 3830,$ min&~&$\sim 3467,$ BAL max.&0.60 \\ \hline
3100, shoulder&0.20&3538, shoulder &~&3110, wpp&0.19 \\ \hline
3014, dip&~&$\sim 3425,$ BAL max.&0.31&2960 --3012, almost plateou&0.37 \\ \hline
3037, peak&0.41&3190, shoulder&~&2860, wpp&0.22 \\ \hline
2997, peak&0.39&3140 shoulder&~&2625, wpp&0.08 \\ \hline
2950, shoulder&0.20&3108 shoulder&~&2540, wpp&0.08 \\ \hline
2865, wpp&0.08&3090 shoulder&~&2195, peak&0.14 \\ \hline
2520, shoulder&0.03&3038, peak&0.44&1080--1750, offscale&1 \\ \hline
2335&0.05&3013, dip&~&937, dip&~ \\ \hline
2210&0.08&2998, peak&0.42&880, peak&~ \\ \hline
1900, dip&~&2985, shoulder&~&870, narrow line&~ \\ \hline
$\sim 1705,$ shoulder&~&$\sim 2950,$ shoulder&0.22&772, shoulder&~ \\ \hline
1570, shoulder&~&$\sim 2850,$ shoulder&0.07&688, dip&~ \\ \hline
1512, dip&~&2210, peak&0.03&620, peak&0.08 \\ \hline
1630--1670 off-scale range&~&1600--1710, off-scale range&1&540, peak&~ \\ \hline
1040--1440 off-scale range&1&1510, dip&~&520, peak&~ \\ \hline
~&~&1080--1440, off-scale range&1&500, peak&~ \\ \hline
990, wpp&~&1170&~&$\sim 480,$ shoulder&~ \\ \hline
840--910 off-scale&~&$\sim 1000,$ dip&~&$\sim 460$ shoulder&~ \\ \hline
870, narrow line, off-scale&~&995, wpp&~&440, dip&~ \\ \hline
790, dip&~&960, shoulder&~&880 (p) $\leftrightarrow $ 997(d)&0.15 \\ \hline
772, peak&0.46&940, dip&~&880(p) $\leftrightarrow $ 688(d)&0.51 \\ \hline
772(p) $\leftrightarrow $ 790(d)&0.30&995(p) $\leftrightarrow $ 1000(d)&0.07&500(p) $\leftrightarrow $ 550(d)& 0.07 \\ \hline
772(p) $\leftrightarrow $ 712(d)& 0.62&995(p) $\leftrightarrow $ 960(d)&0.21&500(p) $\leftrightarrow $ 440(d)&0.43 \\ \hline
750, shoulder&0.03&890--910, off-scale&~&~& \\ \hline
685, peak&0.05&870, narrow line&~&~& \\ \hline
660, peak&0.03&850--867, off scale&~&408, peak&~ \\ \hline
620, peak&0.56&830, shoulder&~&~&~ \\ \hline
607, shoulder&0.23&810, shoulder 790, dip&~&~& \\ \hline
577, wpp&0.09&772, peak&~&& \\ \hline
557, shoulder&0.14&$\sim 752,$ shoulder&~&& \\ \hline
537&0.55&772(p) $\leftrightarrow $ 790(d)&0.23&& \\ \hline
517&0.61&772(p) $\leftrightarrow $ 712(d)&0.54&& \\ \hline
495&0.77&750(s) $\leftrightarrow $ 712(d)&0.29&& \\ \hline
456, dip&~&712, dip&~&& \\ \hline
456(d) $\leftrightarrow $ 495&0.96&685, peak&~&& \\ \hline
443, peak&~&665, shoulder&~&& \\ \hline
423, peak&~&647, dip&~&& \\ \hline
423(d) $\leftrightarrow $ 443&0.09&685(p) $\leftrightarrow $ 704(d)&0.04&& \\ \hline
~&&685(p) $\leftrightarrow $ 647(d)&0.14&& \\ \hline
~&&665(s) $\leftrightarrow $ 647(d)&0.07&& \\ \hline
~&&618, peak&~&& \\ \hline
~&&$\sim 607,$ shoulder&~&& \\ \hline
~&~&588, dip&~&& \\ \hline
~&~&618(p) $\leftrightarrow $ 647(d)&0.57&& \\ \hline
&&618(p) $\leftrightarrow $ 588(d)&0.49&& \\ \hline
&&607(s) $\leftrightarrow $ 588(d)&0.23&& \\ \hline
&&575-580, wpp&~&& \\ \hline
&&555, shoulder&~&& \\ \hline
&&537, peak&~&& \\ \hline
&&517, peak&~&& \\ \hline
&~&557, shoulder&~&~& \\ \hline
&~&497, peak&~&~& \\ \hline
&~&457, dip&~&~& \\ \hline
&~&$\sim 440-450,$ peak&~&~& \\ \hline
&~&417, peak&~&~& \\ \hline&~&424, dip&~&~& \\ \hline
&~&407, dip&~&~& \\ \hline
&~&497(p) $\leftrightarrow $ 577(p)&0.63&~& \\ \hline
&~&497(p) $\leftrightarrow $ 557(p)&0.55&~& \\ \hline
&~&497(p) $\leftrightarrow $ 537(p)&0.18&~& \\ \hline
&~&497(p) $\leftrightarrow $ 517(p)&0.11&~& \\ \hline
&~&497(p) $\leftrightarrow $ 588(d)&0.68&~& \\ \hline
&~&497(p) $\leftrightarrow $ 457(d)&0.97&~& \\ \hline
&~&437(p) $\leftrightarrow $ 457(d)&0.08&~& \\ \hline
&~&437(p) $\leftrightarrow $ 424(d)&0.11&~& \\ \hline
&~&417(p) $\leftrightarrow $ 424(d)&0.03&~& \\ \hline
&~&417(p) $\leftrightarrow $ 407(d)&0.08&~& \\ \hline
\end{longtable}
\end{widetext}

\subsubsection{Grinded samples}
The off-scale was absent in grinded samples and a number of additional 
bands were resolved. Almost the same absorption structure was observed both 
in oriented and unoriented samples after grinding, compare for instance 
peak and dip positions in B-oriented and unoriented samples, Table~\ref{table2}.

\begin{widetext}
\begin{longtable}
{||p{112pt}|p{45pt}||p{112pt}|p{45pt}||p{112pt}|p{45pt}||}
\caption{\label{table2}Spectral positions and amplitudes of the lines registered in grinded 
carbynoid samples and compressed into pellet with KBr. Amplitude was 
determined relatively the peak at 1068 cm$^{-1}$ in A-samples and relatively 
the peaks at 1080 cm$^{-1}$ 1085 cm$^{-1}$ in B- oriented and unoriented 
samples correspondingly.} \\ \hline 
\multicolumn{2}{||c||}{\textit{Unoriented B-sample}}&\multicolumn{2}{c||}{\textit{Oriented B-sample}}&\multicolumn{2}{c||}{\textit{Oriented A-sample}}  \\ \hline
Frequency, cm$^{-1}$& Amplitude& Frequency, cm$^{-1}$& Amplitude& Frequency, cm$^{-1}$& Amplitude \\ \hline
$\sim 3450,$ BAL max.& 0.39& $\sim 3452,$ BAL max.& 0.80& $\sim 3450,$ BAL max.& 0.78 \\ \hline
3290, shoulder& ~& 3290, shoulder& ~& ~& ~ \\ \hline 
3100, wpp& ~& ~& ~& ~& ~ \\ \hline
3032, peak& ~& 3033, peak& ~& 3027, peak& 0.06 \\ \hline
2963, shoulder& ~& 2970, peak& ~& ~& ~ \\ \hline
2942, peak& 0.27& 2933, peak& 0.22& 2940, peak& 0.34 \\ \hline
2895, dip& ~& ~& ~& 2910, shoulder& 0.23 \\ \hline 
2863, peak& 0.14& 2860, shoulder & 0.13& 2863, peak& 0.25 \\ \hline
1750, shoulder& 0.35& 1750, shoulder& 0.22& 1730, peak& 0.75 \\ \hline
1650, peak&0.46& 1650, peak& 0.56& 1630, peak& 1.14 \\ \hline
1465, shoulder& 0.42& 1470, shoulder& 0.18& 1470, shoulder& 0.57 \\ \hline
1438, shoulder& 0.62& 1440, shoulder& 0.31& 1440, shoulder& 0.66 \\ \hline
1410, peak& 0.98& 1410, peak& 0.89& 1410, peak& 0.88 \\ \hline
~& ~& ~& ~& 1346, peak& 0.66 \\ \hline
1283, peak & 0.79& 1287, peak& 0.67& ~& ~ \\ \hline
1230-1260, sh-r& ~& 1250, shoulder& 0.71& ~& ~ \\ \hline
1202, peak & 1.14& 1200, peak& 0.93& 1198, peak& 1 \\ \hline
~& ~& 1170, shoulder& 0.91& 1167, peak& 1 \\ \hline
~& ~& 1120, shoulder& 0.84& ~& ~ \\ \hline
1085, peak& 1& 1080, peak& 1& 1068, peak& 1 \\ \hline
$\sim 1040$ wp sh-r& ~& 1040, shoulder& 0.80& 1035, shoulder& 0.80 \\ \hline
950 wp sh-r& ~& 950 wp sh-r& ~& ~& ~ \\ \hline
888, peak& 0.83& 888, peak& 0.60& 888, peak& 0.34 \\ \hline
~& ~& ~& ~& 870, peak& 0.13 \\ \hline
848, peak& 0.60& 848, peak& 0.40& 850, peak& 0.22 \\ \hline
~& ~& 832 (wp sh-r)& ~& ~& ~ \\ \hline
~& ~& 811, wpp& ~& ~& ~ \\ \hline
750, peak& 0.12& 750, wpp (?)& ~& ~& ~ \\ \hline
~& ~& 687, wpp (?)& ~& ~& ~ \\ \hline
$\sim 620$ wpp& ~& 623, peak& 0.11& ~& ~ \\ \hline
605, peak& 0.12& 607, peak & 0.24& 612, peak& 0.17 \\ \hline
570, shoulder& ~& 565 wpp& ~& ~& ~ \\ \hline
516, peak& 0.37& 515, wpp & 0.44& ~& ~ \\ \hline
525, wp sh-r (?)& ~& ~& ~& ~& ~ \\ \hline
492, shoulder& ~& ~& ~& ~& ~ \\ \hline 
478, peak& 0.46& 493, wpp & 0.49& 494, peak& 0.36 \\ \hline
445, wp sh-r (?)& ~& 478, wpp & 0.47& 463, shoulder& 0.21 \\ \hline
455, shoulder& 0.48& 437, dip& ~& ~& ~ \\ \hline
420, dip& ~& 427, wpp& ~& ~& ~ \\ \hline
408, peak& 0.06& 415, dip& ~& ~& ~ \\ \hline
400, dip& ~& 408, peak& 0.22& 408, peak& 0.09 \\ \hline
\end{longtable}
\end{widetext}

Most of the peaks in oriented and unoriented grinded B-samples were coinciding in their positions. To visible differences between oriented and unoriented grinded B-samples should be referred the very pronounced shoulders at 1040 and 1120 cm$^{-1}$ in oriented sample in comparison with those ones in unoriented sample, shift of 1085 cm$^{-1}$ - band in unoriented sample to 1080 cm$^{-1}$ in oriented sample and the presence of a dip at 2895 cm$^{-1}$ the only in unoriented sample. BAL-maximum is observed in oriented grinded B-sample at 3452 cm$^{-1}$ and at almost the same position in 3450 cm$^{-1}$ in unoriented sample. Linewidth $\Delta \nu _{BAL}$ = 355 
$\pm $ 5 cm$^{-1}$ in oriented grinded B-sample. Along with BAL there are also in oriented grinded B-sample peak at 2933 cm$^{-1}$, weak shoulder at 
2970 cm$^{-1}$, peak at 2860 cm$^{-1}$, shoulder at 1750 cm$^{-1}$, peaks at 
1650 cm$^{-1}$, 1410 cm$^{-1}$, 1282 cm$^{-1}$, shoulder at 1250 cm$^{-1}$, 
peak at 1200 cm$^{-1}$, shoulders at 1110 cm$^{-1}$, 1040 cm$^{-1}$, peak at 
1080 cm$^{-1}$, shoulder at 1000 cm$^{-1}$, peaks at 888 cm$^{-1}$, 848 
cm$^{-1}$, then the lines with rather good resolution at 778, 675, 620 
cm$^{-1}$, shoulder at 580 cm$^{-1}$, peaks at 537, 517, 497, 440 (weak), 
415 (weak) cm$^{-1}$. In grinded A-sample BAL has maximum at $\sim $3450 
cm$^{-1}$, that is, in distinction from undamaged A- and B-films, practically at the same value with that one observed in grinded oriented and unoriented B-samples. Linewidth of BAL in oriented grinded A-sample has the value $\Delta \nu _{BAL}$ = 365 $\pm $ 5 cm$^{-1})$, which is, like to positions of maximum, is coinciding within accuracy of measurements with linewidth value of BAL in grinded B-sample. Further, there 
are in oriented grinded A-sample the bands with peak positions at 2940, 2860 cm$^{-1}$, shoulder at 1730, 
peaks at 1630, 1410 cm$^{-1}$, absorption with almost plateau shape in the 
range 1060 --1200 cm$^{-1}$, peaks at 888, 850 cm$^{-1}$, peak at 620 
cm$^{-1}$, shoulders at 540 cm$^{-1}$, 520 cm$^{-1}$, peak at 500 cm$^{-1}$. It seems to be interesting, that
among the lines, which are observed in grinded A- and B-samples there are 
two groups. There is the group of lines which have practically identical 
positions for both the samples (most intensive among them are: $\sim $3450, 
$\sim $2860, 1470, 1440, 1410, 1200, 888, 848 cm$^{-1})$. There is also the 
second group of the lines undergoing the red shift in B- to A-sample sequence. To second group belong the lines with absorption maximum positions at 1080, 1650, 1750 cm$^{-1}$ in grinded 
oriented B-sample. They are observed at 1068, 1630, 1730 
cm$^{-1}$ in grinded A-sample. It is essential that relative value of red shift is approximately the same for each ot these three lines:
($\Delta \nu $/$\nu {\rm } \quad \approx $ 0.012). One should consider the same relative value of red shift as a strong indication that all three lines belong to the same vibration center. Note that there is the line near 2200 cm$^{-1}$ in undamaged films, which also undergoes red shift. It has the position at 2210 cm$^{-1}$ in B-film, its position in A-film is 2195 cm$^{-1}$. The relative value of red shift for this line is $\Delta \nu $/$\nu {\rm }$ $\approx $ 0.007. Hence it seems to be evident that the line near 2200 cm$^{-1}$ should belong to another vibration center. The multicomponent low frequency line with the first rotation peak 
position at 495 cm$^{-1}$ undergoes some violet shift ${\rm }\Delta \nu $/$
\nu $ =0.010). So we have at the minimum three different vibration system in the samples studied.
The next essential feature of absorption character in the samples studied is 
broadening of all the lines in A-samples in comparison with B-samples, Table~\ref{table1},Table~\ref{table2} (compare also Fig.\ref{fig2} with Fig.\ref{fig3}). The spectra, which are presented in Fig.\ref{fig2} and 
Fig.\ref{fig3} were registered on the same IR-spectrometer with the same spectral 
resolution. For instance the width of well resolved b-line at 1410 cm$^{-1}$ 
is 56 cm$^{-1}$ and 85 cm$^{-1}$ in B- and A-samples correspondingly. The 
widths $\Delta \nu _{a}$ for a-line (its maximum frequency position $\nu 
_{a}$ = 1650 cm$^{-1}$ in B-sample and $\nu _{a}$ = 1630 cm$^{-1}$ in 
A-sample) are equaled to $\sim $125 cm$^{-1}$ and $\sim $185 cm$^{-1}$ 
correspondingly. 
The substantial feature of IR-absorption is also growth and drastic 
redistribution of intensities of red shifted lines in the samples A in 
comparison with the samples B. It is convenient to use as a reference 
amplitude the amplitude of any line from the group of two lines 888 and 848 
cm$^{-1}$, amplitude ratio of which is $\sim $ 3 : 2 and remains the same in 
both the series of samples. It was found that amplitudes of red shifted 
lines are increasing relatively reference line at 888 cm$^{-1}$ and gain 
values are $\sim $ 6, 3.6, 1.8 for lines at 1750, 1650, 1080 cm$^{-1}$ correspondingly. 
The relative amplitude ratios in this set are 1: 2.5: 4.5 in B-sample and 1: 
1.5: 1.1 in A-sample. It is evident that \textit{the amplitude alignment takes place in A sample for this set.} This result is additional prove 
that\textit{ all three lines belong to the same vibration structure }that is to the only one type of vibration centers. This conclusion is in 
agreement with conclusion obtained early in IR studies of carbynes, where 
the lines at 1060 cm$^{-1}$, 1600 cm$^{-1}$, 1720 cm$^{-1}$ (1720 cm$^{-1}$ line 
is mostly appeared as a shoulder) \cite{Evsyukov1999, Sladkov1989}, (that is 
the lines with the frequency positions near those ones of red shifted three lines) 
 have been observed and were referred along with line near 2200 
cm$^{-1}$ to carbyne localized vibration modes characterizing own carbyne 
structure (characteristic carbyne lines). It should be noted that BAL was 
not interpreted in previous studies and correspondingly not referred to 
characteristic carbyne lines.
The lines in spectral range (920 - 1850) cm$^{-1}$ (A-sample), (940 - 
1820) cm$^{-1}$ (B-sample) as well as in the range $\sim $(400 - 700) 
cm$^{-1}$ seem to be superimposed with background lines: MB-line and LB 
-line respectively. The conclusion about the spectra imposition is very 
similar to analogous conclusions of many authors. So Gerasimenko et al \cite{Gerasimenko1978} 
observed in Si implanted with 2.5 MeV hydrogen ions in addition to peaks 
that are clearly resolved the considerable unresolved background absorption 
in the range of $\sim $1800 to 2100 cm$^{-1}$. MB-line reveals the red 
shift of its maximal amplitude position $\nu _{MB}$ along with the 
broadening in B- to A-sample sequence like to above described 
characteristic localized modes. The maximum position of MB-line 
$\nu _{MB}$ in B-sample is observed approximately at $\sim $1100 cm$^{-1}$ 
and linewidth $\Delta \nu _{MB}$ is $\sim $330 cm$^{-1}$. In A-sample 
the maximum position $\nu _{MB}$ is approximately at $\sim $1070 cm$^{-1}$ 
and $\Delta \nu _{MB}$ is $\sim $420 cm$^{-1}$. There is also the 
correlation of intensity of MB-line with the impurity increase. However the 
effective rate of the increase of intensity of MB-line with impurity 
increase being equal to $\sim $1.9 is coinciding with that one of BAL but is 
lower in comparison with the rate of intensity increase of red shifted lines 
at 1650 and 1750 cm$^{-1}$, registered in the same spectral region.
Note that along with unshifted and red shifted lines some new lines have 
been observed in sample A and some lines have disappeared. So there is 
shoulder at 2910 cm$^{-1}$ together with main line at 2940 cm$^{-1}$ instead 
of a single line at 2933 cm$^{-1}$ and peaks at 1346, 870 cm$^{-1}$. At 
the same time peak at 1287 cm$^{-1}$, shoulders at 1120, 1250, peak at 515 
cm$^{-1}$ have not been observed. Instead of two peaks at 3033, 3020 
cm$^{-1}$ and instead of two peaks at 623, 607 cm$^{-1}$ the only 
single-peak bands with intermediate frequencies at 3027 and 612 cm$^{-1}$ 
correspondingly have been observed.

Therefore IR-spectra of carbynoid samples are rather rich and complicated 
and a number of their peculiarities (for instance detailed structure and 
shape of band near 3000 cm$^{-1})$ are subject of additional study. The most 
interesting at present study seem to be spectral characteristics of thet lines which have the relevance immediately to carbyne structure. As 
seen, they undergo red shift, intensity redistribution in own set by total 
intensity increase and broadening (along with all resolved lines) in A- to 
B-sample sequence. These properties are correlating with the total impurity 
content increase. Certain interest represents the vibration-rotation band, 
registered in dispersion mode and observed both in carbynoids and starting 
PVDF-samples. 

\subsection{IR studies in PVDF-samples}

IR spectra of starting PVDF film samples both unoriented and uniaxially 
oriented are characterized by quite flat background line in the range 1800 
-- 3600 cm$^{-1}$ for both type of PVDF-samples, Fig.\ref{fig4}. It means as was 
mentioned above that really broad asymmetric line with maximum near 3450 
cm$^{-1}$ is characteristic line the only for carbynoids (or carbynes). 
(This line cannot belong to possible uncontrolled substances in KBr since it 
is presenting in both ungrinded carbynoid films and grinded carbynoid 
samples compressed into pellet). At the same time the band with two peaks 
near 3000 cm$^{-1}$ (at 2992, 3033 cm$^{-1})$ and dip at 3013 cm$^{-1}$ in 
oriented sample and two peaks at 2990, 3034 cm$^{-1}$, dip at 3013 cm$^{-1}$ 
in unoriented sample is very good pronounced. This band was to some extent 
discussed in Sec.A. Additionally in unoriented PVDF-sample the peaks at 
1700, 772, 620, 538, 516, 495, 439, 414 cm$^{-1}$ were observed. The range 
800--1440 cm$^{-1}$ due to off-scale is the only partly informative and 
the positions of only some dips can be pointed out. Namely, dipshoulders at 
1462, 1330 cm$^{-1}$, dips at 1366, 1000, and 933 cm$^{-1}$ and weak peaks 
at 990, 958 cm$^{-1}$ were found. The spectral lines in oriented PVDF 
-sample were registered almost at the same or even strictly at the same 
positions when comparing with unoriented PVDF-sample. So the peaks at 1708, 772, 618, 537, 514, 494, 438, 415 
cm$^{-1}$, weak peak at 989 cm$^{-1}$, shoulders at 960, 1462 cm$^{-1}$, 
dips at 932, 1000, 1366 cm$^{-1}$, dipshoulders at 960, 1330 cm$^{-1}$ have been observed. It 
is interesting that any absorption bands are absent in the range of 1490 
- 1680 as well as in the range 1730 - 1750 cm$^{-1}$ in both unoriented and 
oriented PVDF-film samples. Consequently these results are direct proof that 
the bands in the ranges 1630 - 1650, 1730 - 1750 cm$^{-1}$ in carbynoid 
samples like to BAL are characteristic bands the only for carbyne structure 
in correspondence with conclusion in Sec.A. 

The oriented PVDF-film was also grinded and pressed into pellet with KBr. 
Unexpected result has been established: the lines which are characteristic 
for carbyne structure were registered, that is broad asymmetric line in the 
range (2300--3700) cm$^{-1}$ with a maximum at $\sim $3480 cm$^{-1}$, 
shoulder at 1750 cm$^{-1}$, and a peak at 1650 cm$^{-1}$. The peaks at 1410, 
1290, 1175, 1090, 886, 848, 475 cm$^{-1}$, shoulders at 1050, 1000, 610, 
515, 450, 435 cm$^{-1}$ were also observed. The relative amplitudes for 
lines at 1750, 1650, 1090 cm$^{-1}$ are approximately 1 : 3: 10, that is, the 
vibration modes at 1750, 1650 cm$^{-1}$ are rather strongly suppressed in 
comparison with the mode at 1090 cm$^{-1}$. Note that relative amplitudes of 
low frequency lines with positions at 515 cm$^{-1}$, 474 cm$^{-1}$ in 
grinded PVDF-sample are essentially stronger than those ones in carbynoid 
A- and B-samples. 

This observation can mean that by grinding a dehydrofluorination takes 
place. In other words the PVDF-sample converts partly to carbynoid 
structure. Analogous phase transition of carbon materials by grinding has 
been reported by Kirda et al \cite{Kirda1979}. They established that the average 
interatomic distance in anthracite sample which has been milled on vibromill 
was decreasing from 145 pm to 128 pm. We consider both the cases to be 
examples of the phase transitions, correspondingly, PVDF $\to $ carbynoid, 
and anthracite $\to $ carbynoid, appearing by the grinding. 

Beside the lines, being to be characteristic for carbyne structure there is 
group of lines in grinded PVDF-sample which have analogues in grinded 
carbynoid samples, at that they have substantially greater relative 
amplitudes in carbynoid structure. Peak positions of lines of this group in 
grinded PVDF are 1410, 1290, 1240 (shoulder), 1175, 1050, 886, 848 
cm$^{-1}$. Their analogues for instance in B-sample are 1410, 1283, 1240, 
band with not pronounced maximum in the range 1230--1260, 1202 
(tentatively), 1040 (shoulder), 888, 846 cm$^{-1}$. Amplitude increase of 
these lines in the samples sequence: grinded PVDF $\to $ grinded carbynoid 
seems to be meaning that indicated lines can be attributed to carbon 
backbone, correspondingly, in grinded PVDF and carbynoid chains and cannot 
be assigned to fluorine, oxygen molecules/molecular ions, their complexes, 
or carbon-fluorine, oxygen-fluorine bonds. 

So, the measurements of IR-absorption in both PVDF-films and grinded 
PVDF-samples are in agreement with conclusions of Sec.A. Furthermore, these 
results can be considered as a direct proof that broad line with maximum 
near 3450 cm$^{-1}$ is characteristic band the only for carbyne structure 
like to the bands in the ranges 1630 - 1650, 1730 - 1750 cm$^{-1}$, 1060 - 
1090 cm$^{-1}$, as well as the band in the range $\sim $ 2100 - 2200, 
which were attributed to characteristic bands for carbyne structure early in 
carbyne studies \cite{Evsyukov1999, Sladkov1989}. The results allow also to suggest that the lines 
with peak positions in grinded PVDF at 1410, 1290, 1240 (shoulder) 1175, 
1050, 886, 848 cm$^{-1}$ can be attributed to vibration modes of carbon 
backbone, that is, it is suggested that they are not responsible for 
possible lateral C-F or C-H valence bond vibrations.

\subsection{Nature of red shift and origin of background \\ IR-lines }

Although the bands in the ranges 1630 - 1650, 1730 - 1750 cm$^{-1}$ as well 
as the band in the range 1060 - 1090 cm$^{-1}$ were attributed to 
characteristic bands for carbyne structure in a number of previous studies 
\cite{Evsyukov1999, Sladkov1989} and references therein, see for instance references [1 - 8, 14 - 29] in \cite{Evsyukov1999}, nevertheless the detailed origin of these bands was not 
established unambiguously. Further, any attention was not drawn on the 
presence of broad background lines LB, MB. The origin of BAL was almost not 
discussed. 

It can be suggested that MB- and LB-lines are the manifestation of the 
interaction between localized IR-modes. This interaction can be realized 
both immediately by localized phonon-phonon interaction and by means of spin 
transfer (and/or charge transfer) through spin-phonon (respectively 
charge-phonon) interaction. In a result any interaction mentioned can lead 
to formation of background lines. Note, that formation mechanism of IR 
LB- and MB-lines seems to be related to formation mechanism of ESR-background 
lines in neutron-irradiated or ion implanted silicon and in electron-, 
neutron-irradiated or ion implanted diamond, see, for instance \cite{Brower1972, Ertchak_Efimov_Stelmakh_1997}. 

\textit{If spin transfer (or joint spin-charge transfer) is determining mechanism, then it has to lead to the broadening of all resolved IR-bands by spin-dopant (both spin- and charge dopants) concentration increase}. \textit{Really all IR-lines are essentially broader in the samples with greater fluorine and oxygen content, compare Figures \ref{fig2} and \ref{fig3}, see also Tables \ref{table1} and \ref{table2}.} It testifies in favour of suggestion. Note that this idea has direct experimental cnfirmation on the case of charge transfer, So broadening of IR-lines was observed by dopant concentration increase due to enhance of charge 
transfer in a number of organic conductors, for example in 
alkali-metal-doped polymeric fullerenes \cite{Winter_Kuzmany_1996}. \textit{The simplest way for realization of spin transfer in the samples studied is the transfer by means of quasiparticles with nonzero spins. Consequently we can suggest that two sets of the localized IR-modes undergoing red shift with the impurity content increase belong to 2 various quasiparticle kinds with nonzero spins. The most probably, that the bands at 1750, 1650, 1080 cm}$^{-1}$\textit{ that is first set modes can be attributed to localized vibration modes of spin-Peierls soliton with ESR-spectrum C-M2, Fig.\ref{fig5}.} (The additional 
arguments see below in Sec.D). 

The spectra presented in Fig.\ref{fig5}, Fig.\ref{fig6} are reproduced from \cite{Ertchak_Carbyne_and_Carbynoid_Structures} and \cite{Ertchak_J_Physics_Condensed_Matter}, but 
they need in short comment. Let us remember that PC C-M1 were attributed in 
\cite{Ertchak_J_Physics_Condensed_Matter} to chemical radicals of unknown origin. The point of view on the origin 
of PC C-M1 can be developed. It is very plausible that C-M1 centers can be 
attributed to private case of chemical radicals, namely to pinned solitons 
which belong to the family introduced by Rice et al \cite{Rice1986}. Essential 
difference of Rice-solitons from SSH-solitons is very strong 
localization of unpaired spins in Rice model. The suggestion that C-M1 
centers can be attributed to pinned Rice solitons seems to be correlating 
with both very pronounced saturation behavior of PC C-M1 by microwave power, 
reported in \cite{Ertchak_J_Physics_Condensed_Matter}, and much smaller g-value deviation from g = 2.0023 for PC 
C-M1 in comparison with that one of PC C-M2. We suggest that the line at 
2210 cm$^{-1}$ can belong to Rice solitons, which are, from other hand, 
responsible for ESR-spectrum C-M1, Fig.\ref{fig5}. The correlation of ratio of 
intensities of ESR lines of PC C-M1 and C-M2 with ratio of intensity of 2210 
cm$^{-1}$ line to total intensity of 1750, 1650, 1080 cm$^{-1}$ lines in 
B-sample and their correlating change in B- to A-sample sequence, see Fig.\ref{fig2}, 
\ref{fig3}, \ref{fig5} can testify in favour of this suggestion. (Note that intensity of ESR 
absorption is determined by means of double integration of resonance 
spectrum).

With regard to PC C-M2, one can see that their ESR-spectra are more 
complicated. In the simplest case PC C-M2 spectrum consist of one group of 
unresolved or partly resolved lines which is observed in both A- and 
B-samples. In accordance with \cite{Ertchak_Carbyne_and_Carbynoid_Structures,Ertchak_J_Physics_Condensed_Matter} PC C-M2 represent themselves 
spin-Peierls solitons. There are additional lines L1, L2, L3 in C-M2 
ESR-spectrum of A-samples, Fig.\ref{fig5}b. They correspond to spin wave resonance 
(SWR) which seems to be associated (see below) with spin-Peierls soliton 
lattice formation. The lines L1, L2, L3 are SWR modes with n = 1, n = 
3, n = 5 correspondingly. It should be noted that by appearance of spin 
waves usual PC C-M2 spectrum is in fact simultaneously the mode with n = 0 
in SWR-spectrum. The L1-line is therefore the line consisting of 2 
overlapped SWR-modes with n = 0 and with n =1. The SWR-modes with n = 0 and 
with n =1 were resolved in C-M2-spectrum of A-sample registered after 1.8 
year storage, owing to splitting of all SWR-modes with n $\ne $ 0 into 2 
components, Fig.\ref{fig6}. One can see from Fig.\ref{fig6}, that L1-line really consist of 3 
components. 

Note that the lengths of SWR modes allow to determine carbon chain 
polyconjugation length. Moreover they in principle \textit{all give directly} the value of 
polyconjugation length. Really it should be $\lambda _{1}$/2, where ${\rm 
}\lambda _{1}$ is the length of the first SWR-mode. The vacuum value 
of $\lambda _{1}$ is 3.19 cm in the experiment presented in Fig.\ref{fig6}. Refraction 
coefficient $n_{X}$ in X-range of microwave frequencies is unknown for 
carbynoids. Suggesting $n_{X}$ in the range 1.6 - 160 we obtain $\lambda 
_{1}$/2 = 1cm - 100${\rm}\mu $m, that is in fact the value up to 
sample size. In particular if $n_{X}$ = 2.28 we obtain the value of 
polyconjugation length strictly equal to the sample size in 0.7 cm. Therefore 
we are dealing in fact with sufficiently homogeneous system of carbon 
chains, which \textit{are long, their length is comparable with sample size. Direct 
evaluation of conjugation length corresponds therefore to initial conception on carbyne structure \cite{Sladkov1989}.}

Note that there is a qualitative similarity in the shape of BAL in IR 
spectra with the shape of ESR-lines of PC C-M2. They are also asymmetric, 
see, Fig.\ref{fig5}, Fig.\ref{fig6} with like to BAL opposite to usually observed asymmetry 
extent.

It should be noted that analogous broad asymmetric IR-band with the same 
maximum position near 3450 cm$^{-1}$ (0.43 eV) was observed in \textit{trans} 
-polyacetylene (in photoexcited sample). It is remarkable that it was 
registered simultaneously (on the same signal gain scale) with a number of strongly IR-active localized 
vibration modes of charged SSH-soliton. Nevertheless it was attributed to 
electronic transition, namely to transition from a electrically charged 
SSH-soliton state in the middle of bandgap to the nearest band of 
quasicontinuum energy levels \cite{Vardeny1983}. The fact that vibrational and electronic 
transitions of SSH-solitons can be displayed on the same scale is very 
striking. Usually, typical IR-oscillator strengths are, for example, for 
local impurity modes, as well known, smaller than those ones for electronic 
transitions by a factor of $\sim $ 10$^{3}$. 

We believe that broad asymmetric IR-bands at 3425 cm$^{-1}$ and at 3467 cm$^{-1}$ in B- and A-films correspondingly, which are also displayed simultaneously with IR-vibration localized 
modes (Figures \ref{fig2}, \ref{fig3}), can be attributed to electric spin transition of 
spin-Peierls solitons from their states in bandgap to the nearest 
electric spin-conductivity band of quasicontinuum energy levels E$_{sc}$. In other words since SPS due to spin-charge separation have zeroth electric charge and, consequently, zeroth electric dipole moment, they can be optically active, if they possess by nonzero electric spin moment. Electric spin moment, according to Dirac relativistic quantum theory, 
has pure imaginary value \cite{Dirac1928}. So we suggest that SPS possess by both electric and magnetic spin moments which can be considered as two components of complex electromagnetic spin vector as a single whole. Then the nature of SPS can be developed.They seem to be consisting of two coupled domain walls in both magnetic and electric GSDW, produced by electromagnetic spin-Peierls transition in its generalized form in $\pi$ - and $\sigma$ -subsystems of carbynoids respectively. In other words SPS are domain walls in dimerized both electric $J_{E}$ - and magnetic $J_{H}$-exchange interaction between kink fragments of carbon chain. Therefore SPS represent themselves a new type of solitons, which can be referred to SSH-soliton class. The energy of BAL-transition gives the energy position of SPS states in bandgap. Therefore the energy position of SPS in bandgap is near $\sim $E$_{sc}$ - 0.43 eV in both A- and B-samples. Note that energies of these states are 
strictly coinciding in grinded samples, but they are slightly different (in 
0.006 eV) in films. The asymmetry of the shape of IR-0.43 eV-band in $t$-PA, 
[which to our opinion seems to be appeared due to essential electron-charge 
conductivity on the transition frequency], was the main argument to refer 
0.43 eV-band to namely electronic transition but not to vibration mode \cite{Vardeny1983}. 
The same argument seems to be taking place in the case of spin-conductivity, 
however in contrary to electron-charge transition the pure electron-spin 
transition (as consequence of the spin-charge separation) corresponds [how 
it can be shown] to the line with opposite (mirror) left to right asymmetry 
character, which is appeared due to high spin-conductivity at transition frequency. 

Therefore there is clear correlation of IR-spectra parameters with impurity 
content which in turn determines the formation of spin carriers, that is 
spin-Peierls solitons (and, possibly, related quasiparticles) and spin 
transfer. The charge of SPS has to be zero, since they are products of 
spin-charge separation but they have accordingly to suggestion the nonzero complex spin moment. Then simultaneous IR and ESR activity can be explained if spin-Peierls solitons possess both magnetic and electric spin moments. 
From direct analogy with {t}-PA follows that appearance of BAL with maximum position near 3450 cm$^{-1}$ in IR spectra can be explained as a result of electric spin-transition from SPS
 states in band gap E$_{sc}$ - 0.43 eV to electric spin-conductivity 
band. Results obtained are also in good agreement with proposal that the 
lines at 1080, 1650, 1750 cm$^{-1}$ undergoing the same value of relative 
red shift with impurity increase are SPS vibration modes.

\subsection{Hamiltonian of vibronic system of carbynoids}

SPS, representing the new type of solitons, \textit{electromagnetic} solitons, were introduced in \cite{Ertchak_Carbyne_and_Carbynoid_Structures,Ertchak_J_Physics_Condensed_Matter} as \textit{magnetic} solitons. Here the notion "magnetic soliton", as well as the notions "electric soliton", "electromagnetic soliton" mean correspondingly the solitons, which are formed in magnetically ordered, ferro(antiferro)electrically ordered or simultaneously magnetically and ferro(antiferro)electrically ordered media. Note, that in separate pure ESR or pure optical experiments SPS behave as magnetic or electric solitons, that is, their dual nature seems to be not becoming apparent. Therefore their properties can be compared with the properties of known magnetic and optical
solitons separately. Electric solitons, determined in such a way, are, to our knowledge, unknown at all at present. SPS in magnetic "hypostasis" belong, as it was 
mentioned above, to SSH-soliton class. At the same time the known magnetic 
solitons belong to sine-Gordon class \cite{Izyumov1988}.
The conclusion relatively the SPS type is in accordance with the structure of 
Hamiltonian (see below) which is proposed for theoretical description of 
interacting electronic and vibration subbsystems in spin-Peierls compounds. 
We cannot use the SSH-model for carbynoids immediately since in the SSH-model the equality in 
``rights'' of the part of total wave function, which depend on spin 
variables is ignored. It restricts naturally the application of SSH-model in 
the case of strong electron-electron interaction, taking place in the 
samples studied (see Sec.E). The tradition way to take into account the role 
of spin variables is indirect. It is realized with symmetry requirement to 
total wave function, which leads to appearance of exchange term, 
characterized by constant J, in Coulomb interaction. Consequently to restore 
the role of spin variables we have to add to SSH-Hamiltonian J-dependent 
terms. The corresponding model is well known t-J model, in which both 
spin-phonon and charge-phonon interaction are taken into account. This model 
was proposed by Terai \cite{Terai1981}. The model, used by Terai, can be developed by 
including of additional terms, taking into consideration the interaction 
between the solitons themselves on the one hand and with impurities on the other 
hand (like to known extension of SSH-model). Then the Hamiltonian is:
\begin{equation}
\label{eq1}
\begin{split}
\raisetag{80pt}
 \hat {H}_{t-J}=-&\sum\limits_{i,\sigma}t[1-\alpha(u_{i+1}-u_i)](\hat{c}_{i,\sigma}^+ \hat{c}_{i+1,\sigma}+\hat{c}_{i+1,\sigma}^+\hat{c}_{i,\sigma})\\ 
 + &\sum\limits_{i = 1}^N J[1-\lambda (u_{i + 1}-u_i)(\hat {S}_i\hat {S}_{i+1}-\frac{1}{4}\hat{n}_i\hat{n}_{i+1})  \\ 
 + &\sum\limits_{i = 1}^N{\frac{p_i^2}{2M}} + \frac{1}{2} \sum\limits^{N}_{i = 1} k(u_{i+1}-u_i )^{2}+ \sum\limits_{j = 1}^N {k'}(\delta \theta _j)^2\\
 + &\sum\limits_{q = 1}^N V_q^{imp} \hat {c}_{q,\sigma}^+ \hat{c}_{q,\sigma }  +\sum\limits_{q = 1}^N V_q^{sol}\hat {c}_{q,\sigma }^+ \hat {c}_{q,\sigma}, 
 \end{split}
\end{equation}
where $t$ is hopping integral, $J$ is exchange constant, which is equal to $ J_{H}$ or $ J_{E}$ in the case of ferromagnetic or ferroelectric ordering respectively, $\hat {c}_{q,\sigma }^+$ and $\hat {c}_{q,\sigma }^-$ are creation and annihilation operators of a particle with spin 
\textit{$\sigma $ }on the \textit{q-th} unit site, $\hat S_{q}$ is spin operator, $\hat n_{q}$ is number particle operator, $u_{q}$ is 
the displacement of the \textit{i-th} atomic unit in the lattice, $p_{q}$ is the momentum 
conjugate to $u_{q}$, $\alpha$ and $\lambda $ are charge-phonon and 
spin-phonon coupling constants respectively, $V_q^{imp}$, $V_q^{sol}$ characterize the 
interaction at \textit{q-th} atomic unit site of the solitons with counter impurity complexes and with the solitons on surrounding chains respectively, $k$ and $k'$ are spring constants, which characterize the vibration along the chain axis and 
perpendicular to the chain axis respectively. 
\begin{figure}
\includegraphics[width=0.5\textwidth]{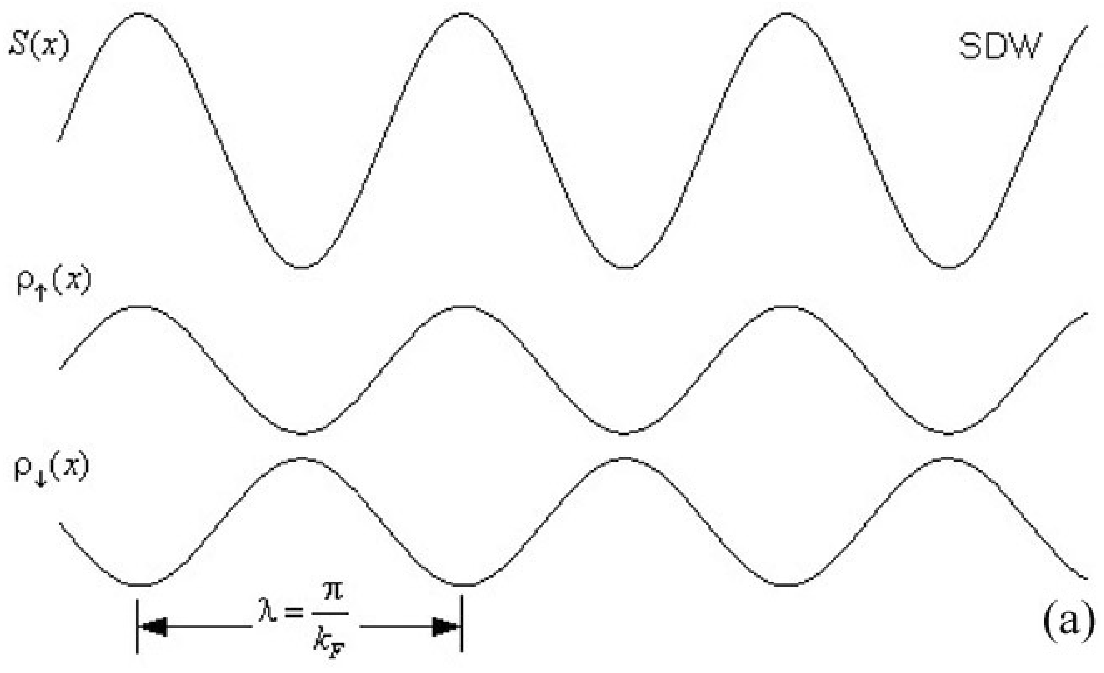}
\includegraphics[width=0.5\textwidth]{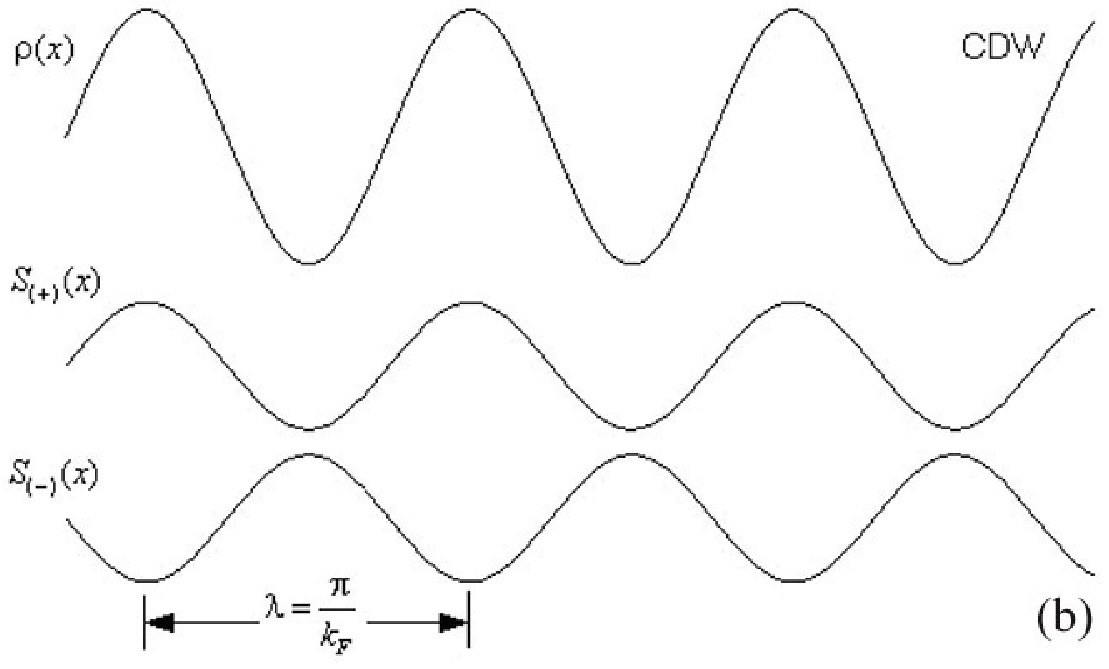}
\includegraphics[width=0.5\textwidth]{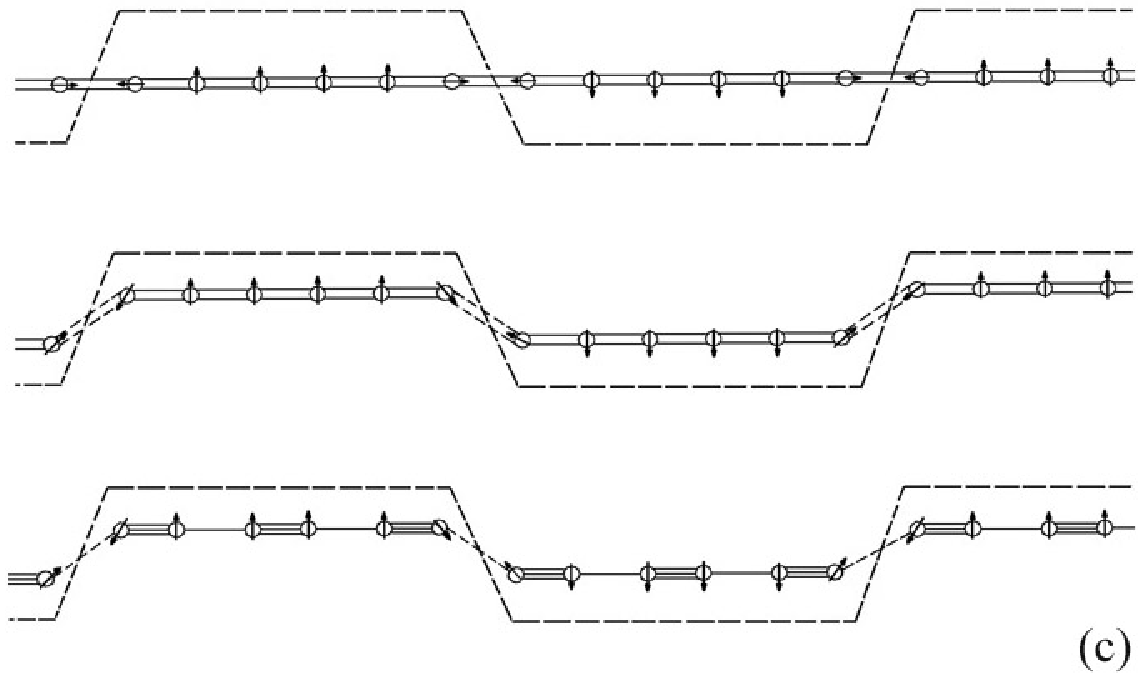}
\caption[Some density wave schemes]{\label{fig7} Some density wave schemes: (a) SDW, here $\rho_\uparrow(x), \rho_\downarrow(x)$ are spatial distributions of charge density corresponding to spin up and spin down subsystems; (b) CDW, here $S_{(+)}(x)$ and $S_{(-)}(x)$ are spatial distributions of spin density corresponding to subsystems with charge densities $\rho (x) - \rho_0 > 0$ and $\rho (x) - \rho_0 < 0$, where $\rho_0$ is average value of charge density before SDW formation; (c) top figure: GSDW formed by generalized spin-Peierls transition, figure in the middle: mixed state with GSDW and kink bond angle alternation (\textit{cis}-configuration) formed in result of simultaneous action of generalized spin-Peierls and generalized Peierls transitions; bottom figure: mixed state with GSDW, kink bond angle alternation (\textit{cis}-configuration) and bond length dimerization in result of simultaneous action of generalized spin-Peierls, generalized Peierls and conventional Peierls transitions.} 
\end{figure}
It can be seen from the expression (\ref{eq1}) (without even solution) that along 
with bond alternation like to SSH-model there is analogous alternation of 
exchange interaction between spin-localizing units. We take for possible the 
predominance in (\ref{eq1}) of $J$-dependent term in the samples studied. Then, \textit{a priori}, it becomes to be evident that the SPS solution will have 
mathematically the form, which is similar to well known soliton solution of 
SSH-Hamiltonian and as a result the appearance of purely magnetic solitons (in the case when $J$ is equal to $ J_{H}$) 
belonging to SSH-family in distinction from the known sine-Gordon magnetic 
solitons. 
\textit{Therefore modified t-J model above proposed seems to be reasonable and necessary generalization of SSH-model for organic conductors in the case of the systems undergoing the spin-Peierls or generalized spin-Peierls transitions, that is for the systems for which the electron-electron interaction is rather strong.} Really it is known that the SSH-model is inapplicable immediately for the 
case when on-site Coulomb interaction U is $>$ 4.68 eV \cite{Subbaswamy_1981}. The fact that 
$t-J $model proposed can be considered in its kernel as a SSH-model generalization 
follows from Hamiltonian (\ref{eq1}) since if $J = 0$ we have the extended SSH-model \cite{Li_Xing_Yao_1991} in 
which the possibility of soliton-lattice formation is taken into account (by 
last term).

\subsection{Evaluation of U-value and comparison with quantum field theory 
predictions }
According to Horovitz \cite{Horovitz1982} the number of IR active modes of topological 
soliton in polymer chain is determined by the number of degrees of freedom of elementary monomer unit 
of the chain. It seems to be the most probably that three above mentioned 
red shifted IR-lines in the samples studied are the modes associated with 
three degrees of freedom which don't bring out the atoms of elementary unit 
from the chain direction. Consequently it can mean that at least $n$ = 4, that 
is 4 carbon atoms belong to the single linear atomic unit, taking part in 
spin-bond alternation process during spin-Peierls transition. In fact the 
number $n$ can be $>$ 4, since, first, not all the modes were resolved well in 
rather complicated IR-spectrum, second, several modes can be strongly 
coupled and be displayed by a single line in a spectrum and, third, not all 
IR active modes can be localized modes \cite{Li_Xing_Yao_1991}. Actually the resolution in 
carbyne characteristic spectral range was achieved in the spectra of third 
set of samples (which are not reported in this work). Along with peak at 
1086 cm$^{-1}$ the peaks at 1655, 1730, 1789 cm$^{-1}$, shoulders at 1620 and 1700 cm$^{-1}$ were 
observed. According to Horovitz criterion the number of carbon atoms in 
dimerization unit in this sample is $\ge $ 7. Note that the present result 
is in fact the additional proof of Heimann et al suggestion on the kinked shape of 
carbyne chain \cite{Heimann1983,Heimann1984}. On other hand it means, that spin-Peierls transition 
should be namely in generalized form discussed in \cite{Ertchak_Carbyne_and_Carbynoid_Structures}, that in its turn 
indicates that electron-electron correlations in carbon chain are really 
strong. Additional argument to this conclusion is very large shift $\Delta 
$E of SPS energetic states in the gap from midgap position, 
which can be simply evaluated if one takes into account that gap value is 
1.86 eV \cite{Ertchak_J_Physics_Condensed_Matter}, consequently $\Delta $E equals to 0.5 eV. The value of shift 
$\Delta $E indicates directly according to \cite{Heeger_1988} that electron-electron 
correlations are rather strong. This result can be considered to be 
experimental proof of necessity of J-term inclusion in equation (\ref{eq1}) for 
description of vibronic states in carbynoids.

Above proposed modified t-J model is equivalent to some extent to Hubbard 
model, for instance at small J values t-J model is strictly equivalent to 
large U-values Hubbard model \cite{Hellberg1993}. Consequently we can use for the 
evaluation of U the approach of Kivelson and Heim \cite{Kivelson1982} and simple expression 
for U derived there: 

\begin{equation}
\label{eq2}
U \sim 2\xi \Delta E,
\end{equation}
where 2$\xi $ is the width of the soliton in relative interatomic distance 
units. For evaluation of U we have used the same value of relative width of 
free spin-Peierls solitons in B-samples equaled to 2$\xi $ = 15 like the 
width of SSH-soliton in $t-$PA. The value of U = 7.5 eV obtained in that way 
agrees in principle with conclusion of Subbaswamy, Grabowski that the ground 
state in organic conductors at large U (at U $>$ 4.69 eV in the case of 
$t$-PA) is the state with SDW, where distribution of spin density is harmonic 
(further it will be designated as HSDW) \cite{Subbaswamy_1981}. Note that ground state of 
samples studied identified to be also spin density wave with periodic but 
nonharmonic density distribution, it is designated GSDW, see Fig.\ref{fig1}. From 
mathematical point of view the difference between these states is minimal 
since GSDW can be represented as superposition of HSDWs, where amplitude of 
main harmonic is strongly exceed the others. Actually the expansion of GSDW (GCDW) profile in Fourier series is:
\begin{widetext}
\begin{equation}
\label{eq3}
f(x') = A(2L_0 + L_1)\sum\limits_{m = 0}^\infty  {\left\{ {\frac{{\left[ {2(2L_0  + L_1 )\left( {\cos \frac{{\pi mL_0 }}{{2L_0  + L_1 }} - \cos \frac{{\pi m(L_0+ L_1)}}{{2L_0+L_1}}} \right) - \pi L_1 m\sin m\pi } \right]}}{{2L_1 \pi ^2 m^2 }} \cdot \cos \frac{{2\pi mx'}}{L}} \right\}}, 
\end{equation}
\begin{equation}
\label{eq4}
f(x) = A(2L_0  + L_1 )\left[ {\frac{1}{4} + (2L_0  + L_1 )\sum\limits_{m = 1}^\infty  {\frac{{\left( {\cos \frac{{\pi mL_0 }}{{2L_0  + L_1 }} - \cos \frac{{\pi m(L_0  + L_1 )}}{{2L_0  + L_1 }}} \right)}}{{2L_1 \pi ^2 m^2 }} \cdot \cos \frac{{2\pi mx}}{L}} } \right].
\end{equation}
\end{widetext}

\begin{figure}[t]
\includegraphics[width=0.45\textwidth]{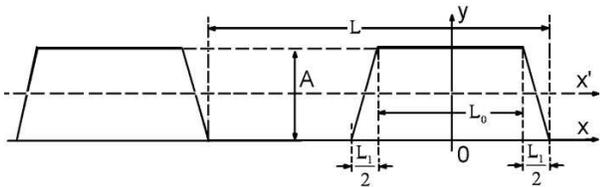}
\caption[SPDW (CPDW) profile along the chain direction]{\label{fig8}SPDW (CPDW) profile along the chain direction,which is formed in result of generalized spin-Peierls (generalized Peierls) transition.}
\end{figure}

The sense of the parameters L$_{0}$, L$_{1}$, L is seen from Fig.\ref{fig8}. The 
functions $f(x)$, $f(x')$ denote spin density distributions $S(x)$, $S(x')$ in the 
case of GSDW and charge density distributions $\rho (x)$, $\rho (x')$ in the 
case of GCDW. Expression (\ref{eq3}) corresponds to position of chain atoms along 
$X'$-direction, Fig.\ref{fig8}, that is like to that one presented in top schema of 
Fig.\ref{fig7}c. Expression (\ref{eq2}) corresponds to position of chain atoms along 
$X$-direction, Fig.\ref{fig8}. We see in this case, that is, by asymmetric deviation of 
spin (or charge) density relatively the chain direction, that the constant 
component is appeared. It can lead in the case of GSDW to ferromagnetic 
ordering if spin distribution is magnetic spin distribution and to 
ferroelectric ordering by electric spin distribution. One can also see from 
(\ref{eq3}) and (\ref{eq4}) that the amplitude of components with m $> 1$ is inversely 
proportional to $m^{2}$, that is, it diminishes rather quickly when number $m$ 
increases. In the first order approximation the relations (\ref{eq3}) and (\ref{eq4}) can 
be represented by
\begin{equation}
\label{eq5}
f(x') = A\,(2L_0  + L_1 )\,\alpha \cos \frac{{2\pi x'}}{L} + o\left( {\frac{1}{{m^2 }}} \right),
\end{equation}
\begin{equation}
\label{eq6}
f(x) = A\,(2L_0  + L_1 )\left( {\frac{1}{4}\, + \frac{1}{2}\alpha \cos \frac{{2\pi x}}{L}} \right) + o\left( {\frac{1}{{m^2 }}} \right),
\end{equation}
where $\alpha $ is
\begin{equation}
\label{eq7}
\alpha  = \frac{{\left[ {(2L_0  + L_1 )\left( {\cos \frac{{\pi L_0 }}{{2L_0  + L_1 }} - \cos \frac{{\pi (L_0  + L_1 )}}{{2L_0  + L_1 }}} \right)} \right]}}{{L_1 \pi ^2 }}.
\end{equation}
Thus GSDW (GCDW) can really be approximated by HSDW (HCDW).

We will see in Sec.F however that the state with conventional SDW in 
B-samples is modified into magnetic spin-Peierls soliton lattice in 
A-samples. Note that according to \cite{Terai1981} in spin-Peierls systems with the 
solitons there is always SDW-order that additionally testifies in favor of 
conclusion on the equivalency of the classifications of the ground state of 
the samples studied as SDW-conductors like to \cite{Ertchak_J_Physics_Condensed_Matter} or generalized spin 
-Peierls conductors like to \cite{Ertchak_Carbyne_and_Carbynoid_Structures} and the model in \cite{Ertchak_Carbyne_and_Carbynoid_Structures} is the development of the model proposed in \cite{Ertchak_J_Physics_Condensed_Matter}.
Our results are gualitatively agreeing with guantum field analysis of 1D 
electronic systems. So the bosonization of 1D-system of fermions with spin 
in the strong coupling limit (that is like the case of the samples studied) 
by using of Jordan-Wigner transformation instead usually used linearization 
procedure leads to the representation of the electronic system which is 
effectively interacting with phonons as a Luttinger liquid, which in its 
turn can be described in terms of two independent scalar fields, namely, the 
fields of CDW holons and SDW spinons \cite{Schmeltzer1994}. Our results testify that 
predicted in \cite{Schmeltzer1994} phenomenon of spin and charge separation is taking place 
even in not strictly 1D-samples, but in quasi-1D-samples. However in 
contrast to \cite{Schmeltzer1994} our experimental data indicate that the excitations which 
carry the separate spin and charge quantum numbers are solitons instead of 
free holons and spinons. In fact our results confirm that the separation 
mechanism of spin and charge in the samples studied has the topological 
nature proposed by Su, Schrieffer, Heeger \cite{Su_Schrieffer_Heeger_1979,Su_Schrieffer_Heeger_1980} and it is unlike to 
deconfining mechanism into free holons and spinons. Note that deconfining 
mechanism is result of mean field approach, which consequently seems to be 
unapplicable in this case. This conclusion is in agreement with the results 
obtained by Mudry and Fradkin \cite{Mudry_Fradkin}. They used instead of mean-field 
approximation the effective quantum field theory which includes the 
fluctuations of order parameters, characterizing the spin liquid. They 
showed that the separate spin and charge excitations of strongly interacting 
electronic system are topological solitons in a Luttinger liquid, that is, 
in other words, the separation of spin and charge in the 1D t-J model is 
determined by topological mechanism in agreement with our data. 

The experimentally observed ordering is also agreed with \cite{Mudry_Fradkin} and can be 
explained in the frame of quantum field model used in \cite{Mudry_Fradkin} as a second order 
phase transition. According to \cite{Mudry_Fradkin} it can be induced from gapless Luttinger 
spin liquid due to frustrations and leads to long-range spatial order, in 
particular to dimerization or other types of order. 

So our results allow to conclude that spin-Peierls-solitons are elementary 
excitations in carbynoid conductors, obtained by dehydrohalogenation of 
polyvinylidene fluoride. We \textit{can conclude on example of carbynoid samples studied that in spin-Peierls systems the role of spin-carriers seems to be predominating in comparison with the role of electric charge carriers by interaction with external electromagnetic field in both the radiofrequency and optical ranges that is both magnetic and electric spin-carriers determine main features in IR- and ESR-spectra.}

\subsection{Spin-Peierls soliton-lattice formation}

The most strong spin-conductivity is observed in the samples of A-group, 
that is in the samples with a maximal impurity content, since the only in 
these samples magnetic spin waves can be excited and detected. It seems to 
be understandable that with spin-Peierls solitons' concentration increase in 
result of spin-dopant (molecules and clusters from residual and 
technological impurities) increase the (quasi)continuous spin medium, which is 
necessary for spin waves' propagation, can be created at some spin-Peierls 
solitons' concentration value. Note that the process of SPS 
creation by doping can be considered to be entirely analogous to the process 
of creation of charged SSH-solitons by electric charge-doping procedure 
\cite{Heeger_1988}, that is, the same type of quasiparticles can be produced by various 
doping species. The spacing between SPS introduced by 
doping is naturally decreased by increasing the dopant concentration and it 
can become to be comparable with the width of a SPS. 
Coherent interaction of SPSs leads then to the spin-Peierls 
soliton lattice (SPSL) formation in a manner similar to the topological 
SSH-soliton lattice formation in \textit{trans}-polyacetylene discussed for instance in 
\cite{Conwell1989, Stafstroem1991}. Accordingly, in bandgap, appearing in result of generalized 
spin-Peierls transition, the narrow magnetic spin-conductive band should be 
additionally formed in the samples A instead of deep individual states of 
SPSs. The presence of spin-conductive narrow band of 
SPS states in forbidden spin-Peierls gap instead of 
isolated spin Peierls-soliton energy levels seems to be necessary condition 
for magnetic spin wave excitation associated with these solitons. It seem to 
be evidently that the appearance of magnetically ordered condensed state 
with SPSL has to lead to broadening of BAL, that is IR-line which now 
correspond to transition of spin carriers \textit{from narrow SPS spin-conductive band in forbidden gap} into main spin-conductive band 
and especially to the broadening of SPS vibration modes. 
Really $\Delta \nu _{BAL}$ changes from $\approx $ 730 cm$^{-1}$ to 
$\approx $ 840 cm$^{-1}$ in A to B film sample sequency (a factor of 1.15), the width 
$\Delta \nu _{a}$ of $ a$-line observed at 1650 before SPSL-formation and at 
1630 cm$^{-1}$ after magnetic SPSL-formation increases from $\sim $125 cm$^{-1}$ to 
$\sim $185 cm$^{-1}$ (a factor of 1.48). It should be noted that the appearance 
of ordering in A-samples of ferromagnetic type \cite{Ertchak_J_Physics_Condensed_Matter} is agreed well with 
lattice formation from SPSs, since they have a shape of SSH 
solitons' type. Physically it means that in distinction from HSDW, where 
spin density distribution is harmonic, and from GSDW, where spin density distribution 
is nonharmonic but symmetric relatively zero line, nonharmonic SPSL is also 
in fact density wave (DW) with also periodic, but nonsymmetric relatively 
zero line spin distribution (zero line position is determined by positions 
of centers of carbon atoms in chain before metal-insulator transition). SPSL 
has the only positive sign in coordinate distribution of spin density while 
in HSDW and in GSDW the positive and negative signs are alternating. Nevertheless 
the term SDW can be kept in principle for DW of these 3 types.
\textit{The observation of Dyson-like shape of ESR-lines, Fig.\ref{fig5}, with the asymmetry of the shape being to be opposite to conventional Dyson effect is [how it can be shown] the direct indication of spin-conductivity effect.} It should be noted that identical asymmetry character of ESR-lines was observed in other spin-Peierls conductors, for instance, in inorganic compound CuGeO$_{3}$ \cite{Palme1996} (see Figure 1 in \cite{Palme1996}, however the authors of \cite{Palme1996} did not present any comments to this result). 
\textit{The conclusion about SPSL formation is also in agreement with calculation in \cite{Li_An_Liu_Yao_1994} of infrared active localized modes for SSH-soliton lattice in {t}-PA, where the authors have found that the frequencies of soliton vibration modes are decreased and their localization is weakened when the dopant content is increasing. Qualitatively similar picture has to be taking place for SPSL since SPSs and SSH-solitons belong mathematically to the same class. The picture predicted is really taking place. It is proved by observation in A-samples, respectively, of red shift and broadening of vibration modes of SPSs.} 
It has to pay attention on the following peculiarity of carbynoid conductors 
studied. \textit{The increase of impurity concentration has to be resulting like to 3D-semiconductors in "spin-metallic" state formation}, that is, in the state without forbidden gap, where spin-conductivity is 
realized in spin-conductivity band ($s.-c$ band), like to charge conductivity in 
usual metals. \textit{However, in correspondence with the measurements of the microwave photoconductivity, the samples studied have a forbidden energy band equaled to 1.86 eV.} Note that analogous situation takes place in heavily ($\sim 
$11{\%}) sodium doped $t$-PA \cite{Yamashiro1997}. \textit{It was concluded in \cite{Yamashiro1997} that soliton lattice formation preserves the band gap presence as well as 1D-character of the sample properties.} Given conclusion was confirmed by 
Hartree-Fock calculations taking into account 3D-interactions due to 
Na-superlattice by using of SSH + extended Hubbard model for so heavily doped 
$t$-PA. Soliton lattice state with a gap is preserved even if quantum 
fluctuations of electron and lattice subsystems were completely taken into 
account. Although in Na-doped $ t$-PA the soliton lattice is formed from 
starting coupling-order wave (private case of CDW), the situation in the 
samples studied seems to be \textit{a priori }qualitatively similar. In carbynoid A-samples 
SPSL seems to be formed from GSDW or more strictly from mixed GSDW - GCDW state, which 
is similar to soliton lattce in Na-doped 
$t$-PA in the sense that it is also the state with breakdown translation 
symmetry. Note that mixed SDW - CDW state is characteristic for the systems 
described by t-J model \cite{Terai1981}. (It is interesting also that t-J model predicts 
a spin-wave-like excitations in so-called high-frequency limit \cite{Jackeli1997}, which 
were really observed in the samples studied, that is, there isn't 
contradiction between the description on t-J model base and SDW 
-representation of ground state). Let us remember that some mixing of CDW 
state in its private case of coupling-order wave follows from the observation of 
weakly pronounced IR band near 2200 cm$^{-1}$ and, probably, ESR 
CM1-centers, which seem to be indication 
on the presence of single-triple bond alternation. For our discussion is substantial that in the case of mixed 
ground state the electrical gap due to CDW and magnetic gap due to SDW (both 
can be appeared in generalized sense) exist (or not) synchronously. It means that we do 
in any case the correct conclusion on magnetic gap existence from the rsults of \textit{microwave photoconductivity} studies in \cite{Ertchak_J_Physics_Condensed_Matter}. 
It should be noted that effective concentration of impurity centers (impurity centers seem to be 
presenting the dopant units in the form of molecules or their complexes but 
not in the form of single impurity atoms) that is effective doping level is 
essentially lower in correspondence with atomic ratios of number of F and O 
atoms to number of C atoms. It is interesting to try to establish which impurities are responsible for the spin-doping effect. At present we can suggest the presence in the interchain space of the following formations from residual and technological impurity counteratoms: oxygen molecules, fluorine clusters, fluorine-oxygen complexes and complexes including C-H bond. The presence of hydroden complexes was discussed. The spin-doping effect can tentatively be associated with molecules or complexes which have nonzero spin like to oxygen molecules and, probably, fluorine-oxygen complexes. The formation of fluorine-oxygen complexes in interchain space seems to be the consequence of the diffusion mobility of their components. High diffusion mobility of impurity atoms, which aren't included in a main carbon backbone, is a general property for 1D-organic polymers \cite{Kirova1997}. Therefore, it is possible that fluorine-oxygen complexes, for instance $FO_4$, can be formed. The spin-doping effect of the complexes with nonzero spin is expected to be analogous to charge-doping effect of well known charged dopants in 1D-organic polymers, for instance, pure halogen ion groups like to $(J_3)^-$ or $(J_5)^-$ \cite{Yoshino1980, Chien1983} and $(O_2)^-$ ions \cite{Mathys1997}.
We evaluate that doping level in our samples is not exceeding in any case 
(but can be comparable) with that one in heavily ($\sim $11{\%}) sodium 
doped $t$-PA studied in \cite{Yamashiro1997}. \textit{Hence follows immediately that the only SPSL formation will preserve the transition to "spin-metallic" state.}
\textit{Therefore the assumption on the formation of magnetic SPSL in A-samples agrees well with ESR, IR-absorption and }\textit{microwave photoconductivity }\textit{data}. 
The electric SPSL has even more so to be formed, that is it can be expected 
in both type of the samples. However it cannot be registered immediately in 
both film and grinded samples studied by means of IR-absorption 
spectroscopy. It cannot be registered in films owing to off-scale in 
necessary spectral range. It cannot be registered in grinded samples because 
of absence of translation invariance in grinded samples, diluted by KBr, since even 
"polycrystalline" structure on the base of electric SPSL 
seems to be not taking place (but relatively magnetic SPSL it can be). This 
conclusion becomes understandable if one takes into account that lattice 
period of electric SPSL is expected to be much more exceeding lattice period 
of magnetic SPSL. It is also understandable, that presence of electric SPSL 
is sufficient for excitation and propagation of electric spin waves, but in 
general is insufficient for emergence of magnetic spin waves. Although 
electric SPSL is difficult to detect by IR-absorption methods in the samples 
studied, nevertheless the magnetic SPSL was experimentally identified.
Additional argument in favour of SPSL model gives evaluation of possible 
relative change of cinematic constant by magnetic SPSL formation from the 
value of red shift. Actually, the changes in IR-spectra by magnetic SPSL 
formation can qualitatively be explained taking into account general 
spectroscopic properties of polymers on the simple model representing the 
translation-invariant chain consisting of $N$ oscillators with eigenfrequency 
$\nu$ and characterized by cinematic $\tau$ and force $k$ coefficients for any pair of 
neighbouring oscillators. The vibration branch assigned to single soliton 
mode by soliton lattice formation can be given then by the expression, which 
is analogous to that one known for polymer chain \cite{Gribov1977}, that is by 

\begin{equation}
\label{eq8}
\nu _m^2  = \{ 1 + 2\tau \cos (\frac{{m\pi }}{N} + 1)\} \{ \nu _0^2  + 2k\cos (\frac{{m\pi }}{N} + 1)\}, 
\end{equation}
where m = 1, 2, 3,..., N.

For large $N-$numbers the frequency shift between \textit{i-th} localized mode of free 
soliton and N-multiply degenerated mode in soliton lattice is satisfying the 
expression

\begin{equation}
\label{eq9}
\nu _{s,i}^2 - \nu _{0,i}^2 = 2\tau _{s,i} \nu _{0,i}^2 + k_{s,i} + 2k_{s,i} \tau _{s,i}, 
\end{equation}
that is the frequency shift has really to be taking place. The effect is 
similar to that one observed always by polymer formation where vibration 
eigenfrequencies are shifted down relative to eigenfrequencies of 
monomer-units, from which given polymer is built, even in the case when free 
monomer structure and structure of corresponding polymer unit are 
coinciding. To evaluate the contribution in the shift due to soliton lattice 
formation we take into account the fact, that frequency shift observed (that 
is, the shift for the origin of which SPSL formation is proposed to be 
responsible) is small in comparison with starting frequency of soliton modes 
before SPSL formation. Then the expression for the shift of \textit{i-th} soliton vibration mode in 
common case, when both the cinematic and the force parameters can be changed, 
will be:
\begin{equation}
\label{eq10}
\delta \nu _{s,i} = \frac{{{\rm{(}}\nu _{0,i}^2 + k_{s,i} {\rm{) }} \delta \tau _{s,i} {\rm{ }}}}{{\nu _{s,i} }} + \frac{{(1 + 2\tau _{s,i} {\rm{) }} \delta k_{s,i} {\rm{ }}}}{{2\nu _{s,i} }},
\end{equation}
where $\delta\tau_{s,i}$ and $\delta $k$_{s,i}$ are the change of 
the cinematic and the force parameters.
Note that the expressions (8) to (10) seem to be valid in the case of the 
formation of soliton lattice of both the types, that is the lattice from the 
solitons like to SSH-solitons in $\textit{t}$-PA and the lattice from spin-Peierls 
solitons identified in the samples studied. In the state with SPSL the 
electric charge distribution and consequently the force coefficients can be 
considered to the first order approach to be unchanged like to those ones in 
conventional SDW-state. Then we have

\begin{equation}
\label{eq11}
\delta \tau _{s,i} = \frac{{\nu _{s,i}  \delta \nu _{s,i} }}{{(\nu _{0,i}^2 + k_{s,i} )}}.
\end{equation}

The cinematic coefficient seems to be decreasing due to effective soliton 
mass increase in the soliton lattice state in comparison with soliton mass 
in free state. Physical explanation of the effective soliton mass increase 
by soliton lattice formation seems to be consisting in the following: to excit the 
motion of one soliton all the solitons in soliton lattice have to be excited 
simultaneously in a soliton train. 
It seems to be reasonable to suggest that to establish the law for the 
change of cinematic constant $\delta \tau _{s,i}$ for \textit{i}-th soliton mode 
one should use the functional dependence \textit{$\tau =\tau(M)$} for the simplest carbon compounds, 
where the carbynoid chain monomer unit is presenting and which has the 
vibration eigenfrequency near one of the eigenfrequencies of spin-Peierls 
-soliton modes. Then the simple variation procedure can be used. The most 
suitable modelling vibration mode seems to be C = C stretching mode of monomer C 
= C unit which is presenting in ethylene or allene, the eigenfrequency of 
which is near 1650 cm$^{-1}$, that is, it practically coincides with the eigenfrequency of main spin 
-Peierls-soliton mode at 1650 cm$^{-1}$ (which is designated as $a$-band, then $i=a$ in eq.(9) to (11).
Taking into account that for modelling C = C unit the dependence $\tau =\tau(M)$ is $\tau = 2/M_{0}$,
where $M_{0}$ is carbon atom mass, and believing 
the mass to be variable, we have the relation 
\begin{equation}
\label{eq.12}
\tau = 2/M.
\end{equation}
Then the variation of mass according to eq.12 corresponding to the determined by eq.11 
variation of $\tau$ allows to evaluate the change of soliton mass by SPSL 
formation. Physically it means, that by formation of soliton lattice, parameter $\tau$ will 
"feel" surrounding. Surrounding can be taken into account if relation $\tau = 2/M_{0}$ 
is preserved but M becomes variable near $M_{0}$ to that extent to satisfy 
the value of $\delta \tau_{s,a}$ for a-band according to eq.11, which in turn is determined by 
experimentally measured $\delta \nu_{s,a}$. So for a-band ($i = 
a$) we will have:

\begin{equation}
\label{eq13}
\left| {\delta \tau _{s,a} } \right| = \left| {\delta M} \right|\frac{2}{{M^2 }}.
\end{equation}

Consequently, neglecting free SPS mass in comparison with the 
SPS mass in SPSL we obtain the evaluation value for the 
SPS mass ${m}_{sL}$ in SPSL:

\begin{equation}
\label{eq14}
m_{sL} = \left| {\delta M} \right| = 
\frac{{M^{\rm{2}} {\rm{ \nu }}_{{\rm{s}}{\rm{,a}}} {\rm{\delta \nu }}_{{\rm{s}}{\rm{,a}}} {\rm{ }}}}{{{\rm{2(\nu }}_{{\rm{0}}{\rm{,a}}}^{\rm{2}}  + {\rm{k}}_{{\rm{s}}{\rm{,a}}} {\rm{)}}}}.
\end{equation}

Using for the force constant k$_{s,a}$ the value in the range between 15.1 
$\times $ 10$^{6}$ cm$^{-2}$ and 68 eV A$^{-2}$ (17.1 $\times $ 10$^{6}$ 
cm$^{-2})$ corresponding to allene \cite{Gribov1976} and carbyne \cite{Rice1986} we obtain for \textit{the effective value of SPS mass in SPSL the evaluation in 0.11 - 0.13 a.m.u}. We 
obtain also the evaluation of $\delta \tau /\tau$ which is $\sim $1{\%}. Note that this value is coinciding with the evaluation value in $\sim $1{\%} predicted by Mozurkewich et al \cite{Mozurkewich1985} for changes in ratio of Youngs' modulus for condensate and underlying lattice by CDW depinning in CDW-conductors. Mozurkewich et al have suggested in their model that the total stiffness of a CDW conductor can be separated into 
contributions from the CDW condensate and the underlying lattice which are 
additive for pinned CDW, while for a fully depinned CDW the total stiffness 
is determined by the stiffness of the underlying lattice. Although this 
model yields elasticity change within the range of experimentally observed 
shifts Bourne and Zettl \cite{Bourne1987} have used for the explanation of elastic 
properties of CDW conductors TaS$_{3}$, NbSe$_{3}$ in the presence of ac, 
and/or dc electric fields the development of the model \cite{Mozurkewich1985} proposed in \cite{Bourne1986} 
which treats the interaction of deformable CDW with a \textit{deformable} underlying lattice. 
Bourne and Zettl have determined for instance the changes in ratio of 
Youngs' modulus for condensate and underlying lattice by CDW depinning in 
TaS$_{3}$, taking place when ac frequency varies in the range 0 - 10 MHz. 
They were equal to $\sim $4 10$^{-3}$ and are explained by the model 
proposed in \cite{Bourne1986} better. At the same time the model of Mozurkewich et al 
\cite{Mozurkewich1985} proposed for interpretation of elastic properties of CDW conductors 
seems to be more appropriate for changes of optical properties in particular 
of the centers being to be topological disturbances in CDW itself and 
becoming to be pinned by CDW pinning. Really from microscopic point of view 
the changes in Youngs' modulus established in \cite{Mozurkewich1985, Bourne1987, Bourne1986} are equivalent to 
the changes in spring constants for the stretching vibration modes of 
topological defects in CDW that is \textit{$\delta$ k/k} is also $\sim$ 1{\%} for the vibration 
modes of these centers in CDW-conductors by depinning. It is substantially 
that optical transitions are realized for the time which is insufficient for 
any deformation of underlying lattice, that is model developed in \cite{Bourne1986} 
becomes coinciding in limit of undeformable underlying lattice with the 
model proposed in \cite{Mozurkewich1985}. The same evaluated values for $\delta \tau/\tau$ in studied 
SDW-conductor (defined in generalized above sense) by SPSL formation and for 
$\delta$ k/k in above mentioned CDW-conductors by depinning seems to be understandable 
if one takes into account that the relative energies of interaction of 
valence electron condensates with an underlying atomic lattice seem to be 
comparable for SDW- and CDW-states and they seem to be also comparable with 
relative energy of \textit{SPS condensate} formation, that is with the energy of \textit{SPSL} 
interaction with the underlying atomic lattice. It is understandable also 
that the appearance of the spin wave propagation in result of SPSL formation 
can lead to some effective change {$\delta$ k}$_{s,i}$ in the force constants however it 
seems to be necessary to take into account in the second order approach. It 
seems to be interesting to consider the phase transition \textit{GSDW} to \textit{SPSL} from energy conservation 
law, that is, to which form transform the energy of vibration, corresponding 
to difference in the frequencies of SPS vibration modes. We suggest that 
this energy transform directly in the energy of exchange interaction of 
electric spins of SPS which necessary has to emerge in condensed state with 
SPSL. If it so we can evaluate the value of exchange interaction from energy conservation law as follows. Since 
exchange interaction is determined by spin \textit{couples} then evaluation value in 
zero approximation for exchange parameter in SPSL is predicted to be equal to
double value of red shift, that is, of order 34 - 40 cm$^{-1}$. This value 
will be equal to splitting parameter in electric spin wave resonance.
\textit{Therefore the reasonable value in $\sim $1{\%} of relative change of cinematic constant $\delta\tau/\tau$ by magnetic 
SPSL formation obtained from observed value of red shift of main vibration 
mode of SPSs is strong additional indication in favour of the idea of SPSL formation in A-samples. So above listed arguments seem to be sufficient to insist that the formation of magnetic SPSL in A-samples is actually takes place.}

\subsection{Origin of absorption structure in range 395 - 950 cm$^{-1}$ in PVDF- and carbynoid B-films}

Along with fine-structured band at 475 - 477 cm$^{-1}$, \textit{registered in dispersion mode} both in PVDF- and 
carbynoid B-films (see Sec.A) there are strong intensive bands, which are 
also observed in PVDF- and carbynoid B-films, \textit{which however are registered in absorption mode}. The shape of two well 
resolved bands is very similar, that can be considered as evidence of the 
same origin of these bands. Both the bands along with peaks have 
characteristic shoulders. Peak positions are 618, 620 cm$^{-1}$ for the 
first band and they have the same value of 772 cm$^{-1}$ for the second band 
accordingly in oriented and unoriented PVDF films. The first band has 
shoulder at 607 cm$^{-1}$, and shoulder appears at 750 ($\pm $10 cm$^{-1}$) in the second band in both oriented and unoriented PVDF films. \textit{Strictly the same values of peak and shoulder positions have been observed in oriented and unoriented carbynoid B-films}. The most probably, that to the structure consisting of these 2 bands one can refer 
the band with peak near 915 cm$^{-1}$ (strict peak position and the shape of 
this band were not established due to partial off-scale and overlapping with 
other bands), which is observed in oriented and unoriented PVDF- and 
carbynoid B-films. There is also indication that the bands above 
described and fine-structured band at 475 - 477 cm$^{-1}$ produce some 
united structure (US). It is almost the same spacing with the value near 
$\sim $150 cm$^{-1}$. It is evidently that this united structure cannot be 
attributed in PVDF to valence vibrations of C-H or C-F lateral bonds since 
all the components of this structure are observed in carbynoids and, that is 
striking, strictly at the same positions and even with slightly increased 
(by 5{\%}) relative amplitude. Therefore US-bands can belong the only to 
some intrinsic disturbances of C-C bonds in PVDF and accordingly the only to 
some intrinsic disturbances \textit{in $\sigma$-bond subsystem }of carbynoids. Note, that usually $\sigma$-bond 
subsystem is not taking into account by determination of electronic 
properties of carbyne chains ($\pi$-electronic approximation is usually 
used). The origin of these disturbances should seek in peculiarity of PVDF 
structure, evidently in alternation of $\sigma$-bonded CF$_{2}$ and CH$_{2}$ 
groups. Although all carbon atoms form single C-C bond between themselves along the chain in this compound we have to take into account that each of two neighboring carbon atoms interact with two different surrounding atoms in lateral bonds. Then become to 
be possible the following 2 main variants for ground state:
\begin{figure}
\includegraphics[width=0.5\textwidth]{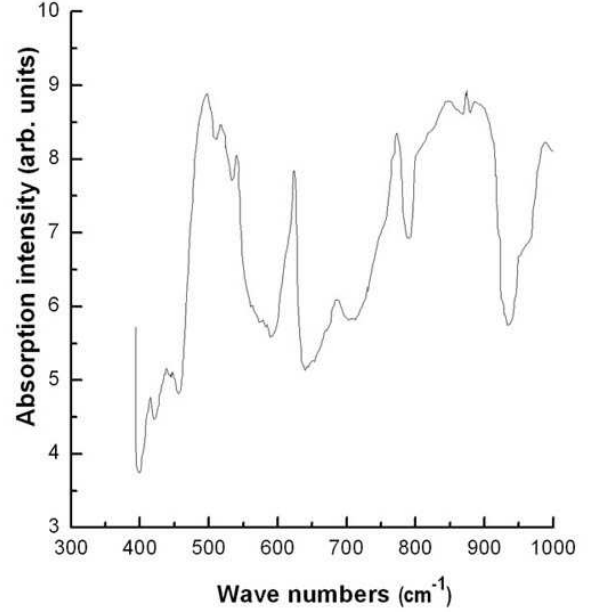}
\caption[Detailed view of spectral distribution of IR absorption intensity in PVDF-film]{\label{fig9} Detailed view of spectral distribution of IR absorption intensity in PVDF-film in the range 395 - 1000 cm$^{-1}$.}
\end{figure}
1) the state with CDW (in the form of HCDW or GCDW) including private case 
of CDW - the state with bond length alternation as a result of Peierls 
transition in generalized or conventional form. Then topological defects in 
corresponding charge distributions will be charged solitons or polarons with 
zero magnetic spin moment.
2) the state with electric spin moment density wave, ESDW, (in the form of EHSDW or EGSDW) 
including private case of ESDW - the state with alternation of exchange interaction or, on bond language, the state with "J-bond" length 
alternation as a result of \textit{electric spin-Peierls} transition in generalized 
or conventional form. Here \textit{we believe in implicit form, that the states with nonzero electric spin have own classification schema}, that is $\sigma$-state relatively magnetic spin value will be most probably $\pi$-state relatively electric spin value. Then topological defects in corresponding electric spin 
density distributions will be solitons or polarons with nonzero electric 
spin.
Naturally the state with conventional magnetic SDW (in the form of HSDW or GSDW) 
cannot be produced due to zero magnetic spin moment in $\sigma$-state of 
macromolecule. In suggestion that case 1 takes place it is 
difficult to explain the appearance of rotational structure in PVDF, since 
C-H and C-F-lateral bonds should take participation in this rotation (due to interaction between charges of carbon backbone and lateral bonds) which 
is evidently unreal in solid matrix. It is also incomprehensible the appearance of 
\textit{dispersion} mode in \textit{absorption} spectra. At the same time case 2 allows to explain all the peculiarities in US-band spectrum observed.
Let us consider this case in some details. Independently of detailed form of 
ESDW we obtain as elementary excitations solitons and polarons (that is, 
soliton and antisoliton pairs) which possess by nonzero electric spin. 
Formation of ESDW corresponds to antiferroelectrically ordered ground 
state. Since the interaction in antiferroelectrics has long range order and 
owing to formation of solitons and polarons the state with continuous 
electric spin distribution in ESDW -form slightly transforms (like to that 
one established in A-samples for magnetic GSDW) into soliton or polaron 
electric spin lattice with substantially more strongly localized electric spins. We 
suggest that electric spin-Peierls polaron lattice (ESPPL) is formed. So we 
 suggest the formation of antiferroelectrically ordered lattice which consist of 2 sublattices A and B. They correspond to soliton and antisoliton components of polaron 
correspondingly. Hamilton operator for ESPPL can be represented in a form:

\begin{equation}
\label{eq15}
\mathcal{\hat H} = \mu_{_B}^{^E}\sum\limits_{\vec {n},\alpha}{\vec {E}}\stackrel{\leftrightarrow}{g}^{^E} \hat{\vec S}_{\vec n}^{^{E(\alpha )}}+\frac{1}{4}\sum\limits_{\vec n \ne \vec m}\sum\limits_{\alpha ,\beta}J_{\alpha ,\beta }^{^{E}}(\vec n - \vec m) \hat{\vec S}_{\vec n}^{^{E(\alpha )}} \hat{\vec S}_{\vec m}^{^{E(\beta )}},
\end{equation}
where $\alpha, \beta = A, B$. Vectors $\vec n, \vec m$ are vectors of the lattice, vector
$\vec E$ is vortex component of total electric intracrystalline and external field, $J_{\alpha ,\beta }^{^{E}}(\vec n-\vec m)$ is exchange integral, $J_{A ,B } < 0, J_{A ,A }=J_{B ,B } > 0$, $\stackrel{\leftrightarrow}{g}^{^E}$ is electrical analogue of magnetic resonance g-tensor, $\hat{\vec S}_{\vec n}^{^{E(\alpha )}}$, $\hat{\vec S}_{\vec m}^{^{E(\beta )}}$ are electric spin operators, corresponding to $\vec n$-element in sublattice A and $\vec m$-element in sublattice B, $\mu_{_B}^{^E}$ is electrical analogue of Bohr magneton. Mathematically similar Hamiltonian is the Hamiltonian for antiferromagnets, which is well known. So we can write its solution right away, following, for example, \cite{Davydov}: 
\begin{gather}
\label{eq116}
\epsilon_1(\vec k) = \frac{1}{2}\left[\alpha(\vec k)-\beta(\vec k)+\sqrt{(\alpha(\vec k)+\beta(\vec k))^2-4\gamma^2(\vec k)}\right],\\
\epsilon_2(\vec k) = \frac{1}{2}\left[\beta(\vec k)-\alpha(\vec k)+\sqrt{(\alpha(\vec k)+\beta(\vec k))^2-4\gamma^2(\vec k)}\right]. \nonumber
\end{gather}
Here $\alpha(\vec k)=\alpha(-\vec k)=L_{AA}(0)-L_{AA}(\vec k)+\mu_{_B}^{^E}E+\gamma(0)$, $\beta(\vec k)=\beta(-\vec k)=L_{AA}(0)-L_{AA}(\vec k)-\mu_{_B}^{^E}E+\gamma(0)$, \, $\gamma(\vec k)=\gamma(-\vec k)=-\frac{1}{2}L_{AB}(\vec k)$, $L_{AB}(\vec k) = S^{^E}\sum\limits_{\vec n \ne 0}J_{\alpha \beta }^{^{E}}(\vec n) e^{i \vec k \vec n}$, $\alpha, \beta = A, B$, $S^{^E}$ is value of the spin. Square root in (16) is arithmetic square root.
Therefore we have two branches of elementary exitation. In long-wave length approximation ($ka << 1$) expressions for $\epsilon_1(\vec k)$ and $\epsilon_2(\vec k)$ can be simplified and we obtain the following dispersion law:
\begin{equation}
\label{eq17}
\epsilon_{1,2}(\vec k) = \pm \mu_{_B}^{^E} {g}^{^E}E + k\sqrt{\gamma(0) (l_{11}+l_{12})},
\end{equation}
where ${g}^{^E}$- tensor is suggested to be isotropic and
\begin{subequations}
\label{eq18}
\begin{gather}
l_{11} = \frac{S}{2}\sum\limits_{\vec n}{\vec n}^2 J_{_{AA} }^{^{E}}(\vec n), \\ 
l_{12} = -\frac{S}{4}\sum\limits_{\vec n}{\vec n}^2 J_{_{AB} }^{^{E}}(\vec n).
\end{gather}
\end{subequations}
Note, that known solution corresponds to continual limit. However, 
in the case of the chain of finite length L the values of $\vec k$ in the first Brillouin zone are discrete. To restore the solution for discrete case one should insert the following boundary conditions: 
\begin{equation}
\label{eq19}
A_{\vec n} = A_{\vec n}+{\vec e}_n L, B_{\vec n} = B_{\vec n}+{\vec e}_n L, \, i.e. e^{i k l} = 1
\end{equation}
Here we have taken into account, that in 1D-case the direction of lattice vector $\vec n$ coincides with chain direction. Then after using of transform to boson operators during derivation of eq.16 we obtain for $|\vec k|$ 
\begin{equation}
\label{eq20}
k_m = \frac{2\pi m}{L}\,
\end{equation}
where m = 0, 1, 2, ... .
Therefore, for the frequences of antiferroelectric resonance we have
\begin{equation}
\label{eq21}
\nu(m) = {g}^{^E}\mu_{_B}^{^E}E + \frac{2 \pi m}{L}\sqrt{\gamma(0) (l_{11}+l_{12})},
\end{equation}
It is seen, that the spectrum of antiferroelectric spin wave resonance (AFESWR) of 1D linear chain with length of L represents in long-wavelength approximation the equidistant set of AFESWR-modes. It is also seen, that this approximation corresponds to linear responce of the system on external electromagnetic field. Then the superposition principle is taking place and one can insist that the mode with $m = 0$, that is fine-structured band at 475 - 477 cm$^{-1}$, is determined by the first part of Hamiltonian, but the modes with $m\ne 0$, that is the bands with peaks at 618, 772 and near 915 cm$^{-1}$ are determined by the second part of Hamiltonian. It is essentially and it allow to explain the difference in registered shapes of $m = 0$ mode and $m\ne 0$ modes in a natural way. Really, in the work \cite{Yearchuck_to_be_published}was shown that optical transition dynamics for the ensemble of two-level optical centers can be described by quantum mechanical optical analogue of Landau-Lifshitz (L-L) equation. 
L-L equation in the case of the ensemble of two-level \textit{noninteracting} optical centers is:
\begin{equation}
\label{eq22}
\frac{\partial \hat {\vec {S}} _{_E} (z)}{\partial t}=\left[ {\hat {\vec {S}} _{_E} (z)\times \gamma_{_E} \vec {E}} \right],
\end{equation}
where vector $\vec {E}$ is:
\begin{equation}
\label{eq23}
\vec {E} = E_{1} e^{i\omega t}\vec {e}_- + E_{1\,} e^{-i\omega t}\vec {e}_+ +\left( {\frac{-\omega_0}{\gamma_{_E}}} \right)\vec {e}_z.
\end{equation}
Two components $E^+ = E_{1} e^{i\omega t}$, $E^- = E_{1\,} e^{-i\omega t}$ are right- and left-rotatory electric 
constituents of external electromagnetic field with amplitude $E_{1}$. The third component $E^z = \left(\frac{-\omega_0}{\gamma_{_E}} \right)$ is 
intracrystalline electric field, which produces the splitting $\hbar \omega 
_0 $ of energy level for each of chain elements into two components. 
Intracrystalline electric field $E_{0}$ corresponds to the frequency $\omega 
_0 $ in accordance with the relation
\begin{equation}
\label{eq24}
\omega _0 \longrightarrow \frac{1}{\hbar }\mu_{_B}^{^E} {g}^{^E} E_0 
= \gamma _{_E} E_0 ,
\end{equation}
where $\gamma _{_E}$ is gyroelectric ratio, that is electrical analogue of gyromagnetic ratio. When proceed in (\ref{eq22}) to the 
observable values and to take into account the relaxation processes, we 
obtain well known optical analogue of Bloch equations. Their solution for 
the case of linear response is 
\begin{equation}
\label{eq25}
S_x^{^E} = \chi_0^{^E} \omega_0 \tau \frac{(\omega_0 - \omega)\tau E_1}{1 + (\omega_0 - \omega)^2 \tau^2},
\end{equation}
\begin{equation}
\label{eq26}
S_y^{^E} = \chi_0^{^E} \omega_0 \tau \frac{E_1}{1 + (\omega_0 - \omega)^2 \tau^2}.
\end{equation}
The complex-valued function of the response, that is complex electrical 
susceptibility, is 
\begin{equation}
\label{eq27}
\chi_{_E} = \chi'_{_E}+ i\chi''_{_E},
\end{equation}
where 
\begin{equation}
\label{eq28}
\chi'_{_E} = \frac{S_x^{^E}}{2E_1}, \, \chi''_{_E} = \frac{S_y^{^E}}{2E_1}
\end{equation}
and 
\begin{equation}
\label{eq29}
\chi_0^{^E} = i |\chi_0^{^E}|
\end{equation}
is pure imaginary quantity, since electric spin vector is a vector with 
imaginary components \cite{Dirac1928}. Then absorption signal, that is $Im \,\chi_{_E}$ 
will be
\begin{equation}
\label{eq30}
Im \, \chi_{_E} \sim \frac{\omega_0 \tau^2 (\omega_0 - \omega) }{1 + (\omega_0 - \omega)^2 \tau^2}\cdot \frac{1}{2}|\chi_0^{^E}|.
\end{equation}
Therefore it really has the dispersion-like shape. Since the signal which is 
determined by the second part of Hamiltonian (15) is proportional to superposition of products 
of $\vec {S}_i \vec {S}_j $, where both the spins are imaginary vectors, it 
will have an absorption-like shape, but it will be registered $\pi$ out of 
phase in comparison with that one detected, for instance, by magnetic 
resonance absorption. Really, the lines registered in US-spectrum in 
absorption mode are $\pi$ out of phase to the signal of dispersion-like 
mode of the same spectrum. This conclusion follows from comparison of the 
signal polarities of observed AFESWR modes with $m \ne 0$ and $m = 0$ in 
reference to the relative polarities of dispersion and absorption signals in 
magnetic resonance, see for instance \cite{Weger_1960}. Splitting energy in 150 cm$^{-1}$ 
is rather large. It means that electric spin-Peierls polaron lattice in C-C 
 $\sigma$-bonds is stable formation and although polaron lattice was emerged 
owing to alternation of CF$_{2}$ and CH$_{2}$ monomers in PVDF-polymer 
structure, it is not needed after its formation in F- and H-atoms for its 
existence. So the observation of the same US-spectrum in carbynoid 
B-samples, which evidently can be attributed to electric spin-Peierls 
polaron lattice in $\sigma$-subsystem, becomes natural explanation and 
confirms in its turn the correctness of conclusion on rather high stability 
of ESPPL being to be formed in $\sigma$-bonds. 
\textit{Therefore, antiferroelectric spin wave resonance being to be optical analogue of antiferromagnetic spin wave resonance has been identified for the first time. Electric spin-Peierls polaron lattice in C-C $\sigma$-bonds is deduced to be responsible for observed AFESWR. The observation of antiferroelectric spin wave resonance (owing to its spectroscopic peculiarities) is the most direct proof of observability of imaginary electric spin moment (at least in condensed matter), which was predicted by Dirac as early as 1928. So, electric spin moment was identified for the first time.}

\subsection{Incommensurate SDW-phase in carbynoids and incommensurate to 
commensurate SDW transition }

Relatively rapid changes in ESR-spectra with storage time in the B-samples 
were reported in \cite{Ertchak_Carbyne_and_Carbynoid_Structures} (however, they were not discussed there). The changes 
observed are determined (how it will be argued below) by the incommensurate 
to commensurate SDW-transition (designation "SDW" is used in generalized sense).
According to \cite{Ertchak_Carbyne_and_Carbynoid_Structures,Ertchak_J_Physics_Condensed_Matter} the SDW-formation in carbynoid films is believed to be 
directly confirming by the observation of a broad ESR-line, which was always 
observed in A-samples by arbitrary sample orientation and at any storage 
time, although broad ESR-line was observable in B-samples the only for 
relatively short time period after their preparation. Broad line is 
attributed to collective ESR-absorption by SDWs themselves. The possibility 
of collective magnetic absorption by SDW means that SDW itself can move 
along underlying chain lattice. SDW-movement can consist in spatial 
variation of their amplitudes and phases and in sliding as it was mentioned 
above. The observed very broad ESR line, Fig.\ref{fig5}, can be atttributed to 
sliding of corresponding SDW, that is, GSDW in our case. In other words it 
can be depinned at the excitation by microwave field, which is used by 
ESR-measurements. This conclusion is in agreement with results described in 
\cite{Bourne1987,Bourne1986,Gruener1985} indicating that characteristic depinning energies for 
CDW-sliding of the known systems correspond to the range of microwave 
frequencies. We bear in mind here that depinning energy for SDW will be of 
the same order with that one for CDW, since SDW ground state can be viewed 
as two CDW states -- one for the spin-up and one for the spin-down 
sub-bands, Fig.\ref{fig7}.
ESR peculiarities, observed in B-samples, Fig.\ref{fig5}, can be explained in the 
following way. Substantial relatively short storage time change in 
ESR-absorption (displaying by the measurements in 12 days) seems to be, 
predominantly, determined by the changes of spin-phonon interaction, being 
in its turn a result of the impurity redistribution. Really, some minimal 
modification of spin-phonon interaction, practically not accompanying by 
atomic rearrangement, can lead to pinning of SDW as well as to spin-Peierls 
soliton pinning, however with different efficiency since wave function of 
SDW-state is delocalized along the chain in distinction from well localized 
domain-wall wave function of the SPSs. The conclusion is in 
agreement with experimentally observed 3.75 times smaller average rate for 
ESR-absorption intensity change by PC C-M2, assigned to spin-Peierls 
solitons in comparison with the rate of intensity change of very broad 
ESR-line, Fig.\ref{fig5}, assigned with SDW-motion. 
Fluorine atom groups, fluorine-oxygen complexes, oxygen molecules and a 
number of complexes with C-H bonds, which are suggested to be formed in 
interchain space, can be redistributed in some time after the sample 
production. There seems to be reasonable to propose that this redistribution 
leads to impurity complexes selfordering, that is, to regularity in their 
positions along chains. Driving force for the selfordering can be, the most 
probably the periodic density wave potential. Residual impurity 
redistribution is the factor, which, in its turn, in self-consistent way can 
affect SDW parameters. The role of impurity redistribution is especially 
effective in density wave pinning if density wave state is near 
incommensurate to commensurate transition. The mechanism is like to the 
mechanism proposed in \cite{Kang1997} and consists in the following. Impurities 
slightly modify the wave vector of SDW (and in principle the lattice 
constant of the underlying lattice) to get the commensurate case. On the 
other hand, when the period of a SDW-modulation becomes commensurate with 
the lattice period, the interaction between two periodicities becomes also 
substantial and the resulting commensurate potential can easy pin the SDW. 
We believe that the case described above takes place in carbynoid B-samples 
and that \textit{the excitation energy for the movement of commensurate GSDW exceeds the energy of X-band microwave photons.} 

Thus the origin of drastic changes in ESR-spectra in relatively short 
storage time presented in Fig.\ref{fig5} can be explained by the redistribution of 
main spin-dopants in the starting periodic SDW-field. SDW-field in turn is 
modified by a new distribution of the impurities. The processes stated seem 
to be self-consistent, and they result finally in the incommensurate to 
commensurate SDW-transition. In its turn the incommensurate to commensurate 
SDW-transition brings to substantial SDW-pinning which results in the 
drastic decrease in ESR-absorption. 
Whether starting SDWs (mixed with CDWs) will be commensurate or 
incommensurate it depends probably on the ratio of carbon atoms to main 
dopant impurity atoms complexes per chain length unity that is probably 
whether is the ratio indicated integer or fractional (the case of the 
ordered impurity distribution between adjacent carbon chains is keeping in 
mind). The concrete chain state, that is with commensurate or incommensurate 
SDWs', will depend on the relative concentration of doping complexes, on 
their distribution function along the carbon chains in interchain space and 
finally on dopant profile after its redistribution during the sample 
storage. The suggestion on the trend to \textit{periodical} selfordering of residual impurity 
complexes in B-samples after their fabrication agrees well with the 
observation of the formation of the \textit{ordered }impurity superlattice in highly Na-doped 
$t$-PA (that is in the material being related to carbynoids) \cite{Zhu_Cox_Fisher_1995,Winokur1987}. 
Consequently, density wave state in carbynoid samples produced by means of 
PVDF dehydrohalogenation can be spontaneously modified in some time after 
the sample production. The modification can lead both to quantitative 
changes consisting in the change of the parameters characterizing the 
SDW-kind, that is depinning energy, wave vector, etc, and to qualitative 
changes consisting in the realization of the incommensurate to commensurate 
transition in density wave state (and \textit{vice versa} in principle). The result stated 
seems to be general for all the density wave conductors and in principle can 
be used to obtain the samples with desirable distribution profile in 
synthetic conductors' technology by means of employing of the periodical 
external fields at the synthesis. 
Note that observed in A-samples broad ESR-line associated with sliding of 
modified SDW (the modification consist in SPS lattice 
formation) as a result of its excitation by microwave field is well 
registered independently on the storage time period. It can be explained if 
to suggest that the SPSL period is incommensurate with atomic lattice period 
of the chain. The possibility of the excitation of SPSL motion or SWR will 
then depend on direction of dc and ac fields relatively chain direction. At 
the same time SPSL in A-samples being to be incommensurate with atomic 
lattice is rather stable formation since it remains incommensurate with 
storage time. It means that the formation of 
SPS lattice stabilizes starting incommensurate SDW-phase and 
the physical properties of the materials. Note that above discussed 
stability properties are concerned the only $\pi$-subsystem of carbynoids 
and will be correct the only in $\pi$-approximation. The results discussed 
in Sec.G show that $\pi$-approximation can be not valid. Especially 
interesting it seems to study the stability of electronic \textit{$\sigma$-subsystem} in both B- and A-samples, 
in particular the stability of ESPPL in B-samples.
\textit{Therefore IR- and ESR-absorption data agree well. They indicate on the formation of rather 
stable SPS lattice in A-samples (that in the samples with rather high impurity content) 
and the formation of unstable incommensurate GSDW state in B-samples (with lower impurity content) 
where the incommensurate to commensurate DW-transition is proposed to be taking place in relatively 
short storage time (12 days) after sample production.}

\section{SUMMARY AND CONCLUSIONS}

The results of IR studies in quasi-1D carbynoid films produced by 
dehydrofluorination of poly(vinilidene fluoride) agree well with ESR-results 
and lead to the conclusion that the carbynoid films produced by 
dehydrofluorination of poly(vinylidene fluoride) are spin-conductors with 
spin-Peierls ground state being stable at room temperature. 

The spin -conductivity increase detected by both IR- and ESR-methods on the 
same carbynoid samples correlates with the concentration increase of 
residual fluorine (and maybe hydrogen, which was uncontrolled) and 
technological oxygen atoms. They tentatively produce in interchain space 
various complexes between themselves. These complexes in their turn can be 
attributed to spin-conductivity dopants. 

The correlation has been established in appearance of red shift of positions 
of IR lines at 1080, 1650, 1750 cm$^{-1}$, position of maximum of broad 
background line (MB - line) in the range of 900 -1900 cm$^{-1}$ (all 
observed in grinded film samples) and 2210 cm$^{-1}$ IR-line (observed in 
starting ungrinded film) as well as in broadening of all observed resolved 
IR-lines with the content of above mentioned residual and technological 
impurities. The lines with the frequencies 1080, 1650, 1750 cm$^{-1}$ 
undergoing the same value of relative red shift (0.012) in the samples with 
more high impurity content (A-samples) are attributed to localized vibration 
modes of solitons of new type, called spin-Peierls solitons. They are 
characterized simultaneously by earlier reported C-M2 ESR-spectrum. 
t has therefore been established that spin-Peierls solitons are simultaneously active, unlike to topological solitons with nonzero spin in \textit{trans}-polyacetylene, in both optical and magnetic resonance spectra. Simultaneous optical and ESR activity of SPS is explained in suggestion that SPS possess by both electric and magnetic spin moments which can be considered as two components of complex electromagnetic spin vector as a single whole. SPS proposed to be consisting of two coupled domain walls in both magnetic and electric GSDW, produced by electromagnetic spin-Peierls transition in its generalized form in $\pi$ - and $\sigma$ -subsystems of carbynoids respectively. In other words SPS are domain walls in dimerized both electric $J_{E}$ - and magnetic $J_{H}$-exchange interaction between short kink fragments of carbon chain and represent themselves a new type of solitons, which can be referred to SSH-soliton class. Real part of electromagnetic spin vector of SPS is proposed to be responsible for magnetic SPS-lattice formation and simultaneous emergence of ferromagnetic-like ordering in A-samples (the samples with the most impurity content of the range used).

IR- and ESR-absorption data indicate that SPSL in 
$\pi$-subsystem of A-samples is incommensurate but rather stable. Starting 
ground state in B-samples (with lower impurity content) is unstable 
incommensurate GSDW state and the incommensurate to 
commensurate DW-transition is proposed to be taking place in relatively 
short storage time (12 days) after sample production. 

The broad asymmetric IR-line with position at the frequency 3450 cm$^{-1}$ 
is attributed to transition of spin carriers from the SPS 
localized levels in bandgap ($E_{sc}$ - 0.43 eV) (in B-samples) or from 
narrow spin-conductivity band in gap, centered near ($E_{sc}$ - 0.43 eV) 
(in A-samples) to main spin-conductivity band. 

Coulomb on-site interaction U is evaluated to be equal 7.5 eV, confirming 
the necessity for the generalization of SSH-model developed for\textit{trans}-polyacetylene. 
The way for the generalization is proposed and it seems to be succesfull in 
the frame of modified t-J model, which in private case of J = 0 coincides 
with the approach based on SSH-model.

The IR-and ESR-data agree well with topological nature of spin-charge 
separation mechanism developed for 1D strongly correlated systems by Mudry 
and Fradkin in the effective quantum field theory consideration which 
includes the fluctuations of order parameters, characterizing the spin 
liquid in contrast to deconfining mechanism of spin-charge separation being 
the result of mean field approach.

Antiferroelectric spin wave resonance being to be optical analogue of 
antiferromagnetic spin wave resonance has been identified for the first 
time. Electric spin-Peierls polaron lattice in C-C $\sigma$-bonds is 
deduced to be responsible for observed AFESWR both in starting PWDF films 
and carbynoid B-films. The observation of antiferroelectric spin wave 
resonance (owing to its spectroscopic peculiarities) is the most direct 
proof of observability of imaginary electric spin moment (at least in 
condensed matter), which was predicted by Dirac as early as 1928. So, 
electric spin moment was identified for the first time.

\begin{acknowledgments}
The authors are thankfull to F.Borovik for the help in the work.
\end{acknowledgments}

\end{document}